\documentclass[11pt, a4paper,prd, nofootinbib, superscriptaddress]{revtex4-1}
\pdfoutput=1 
\usepackage{amsmath,amssymb,amsfonts,dsfont}
\usepackage{lipsum}
\usepackage{bm}
\usepackage{mathrsfs}  
\usepackage{slashed}
\usepackage{feynmp}
\usepackage{graphicx}
\usepackage{amsfonts}
\usepackage{color}
\usepackage{cancel}
\usepackage{amsmath}
\usepackage{makecell}
\usepackage{hyperref}

\usepackage[noindentafter]{titlesec}

\usepackage{siunitx}
\definecolor{orange}{cmyk}{0,0.5,1,0}
\definecolor{rossoCP3}{cmyk}{0,.88,.77,.40}
\definecolor{graa}{rgb}{0.8,0.8,0.8}
\definecolor{blaa}{rgb}{0.2,0.2,0.6}
\hypersetup{
	colorlinks, 
	bookmarksopen, 
	bookmarksnumbered,
	citecolor=blaa, 		
	linkcolor=rossoCP3,	
	urlcolor=rossoCP3,			
}

\def\Tr{\mbox{Tr}\,}

\def\hbar{\hspace{0pt}\raisebox{1pt}{$-$} \hspace{-7pt} h}

\def\5{\overline 5}

\definecolor{JJ}{RGB}{0,144,255}

\newcommand{\be}{\begin{equation}}
\newcommand{\ee}{\end{equation}}
\newcommand{\bea}{\begin{eqnarray}}
\newcommand{\eea}{\end{eqnarray}}

\newcommand{\ba}{\begin{eqnarray}}
\newcommand{\ea}{\end{eqnarray}}

\newcommand{\andeq}{\quad \mathrm{and} \quad}
\newcommand{\LL}{\mathrm{L}}
\newcommand{\RR}{\mathrm{R}}
\newcommand{\dd}{\mathrm{d}}
\newcommand{\transpose}{^{\mathrm{T}}}
\newcommand{\WZW}{\mathrm{WZW}}
\newcommand{\TC}{\mathrm{TC}}
\newcommand{\SU}{\mathrm{SU}}
\newcommand{\Sp}{\mathrm{Sp}}
\newcommand{\U}{\mathrm{U}}
\newcommand{\hc}{\; + \; \mathrm{h.c.} \;}

\newcommand{\EM}{\mathrm{em}}
\definecolor{rossoCP3}{cmyk}{0,.88,.77,.40}
\definecolor{graa}{rgb}{0.8,0.8,0.8}
\definecolor{blaa}{rgb}{0.2,0.2,0.6}

\begin{document}
\title{\textcolor{rossoCP3}{Uncovering new strong dynamics via topological interactions at the 100 TeV collider}}

\author{Emiliano Molinaro}
\email{molinaro@cp3-origins.net}
\affiliation{CP$^{3}$-Origins, University of Southern Denmark, Campusvej 55, DK-5230 Odense M, Denmark,}
\author{Francesco Sannino}
\email{sannino@cp3-origins.net}
\affiliation{CP$^{3}$-Origins, University of Southern Denmark, Campusvej 55, DK-5230 Odense M, Denmark,}
\affiliation{CERN Theoretical Physics Department, Geneva, Switzerland, } 
\affiliation{Danish IAS, University of Southern Denmark, Campusvej 55, DK-5230 Odense M, Denmark,}
\author{Anders Eller Thomsen}
\email{aethomsen@cp3.sdu.dk}
\affiliation{CP$^{3}$-Origins, University of Southern Denmark, Campusvej 55, DK-5230 Odense M, Denmark,}
\author{Natascia Vignaroli}
\email{vignaroli@cp3-origins.net}
\affiliation{CP$^{3}$-Origins, University of Southern Denmark, Campusvej 55, DK-5230 Odense M, Denmark,}
\affiliation{INFN Sezione di Padova, via Marzolo 8, 35131 Padova, Italy.}








 \begin{abstract}
In models of composite Higgs dynamics new composite pseudoscalars can interact with the Higgs and electroweak gauge bosons via anomalous interactions,  stemming from the topological sector of the underlying theory. We show that a future 100 TeV proton-proton collider (FCC-pp) will be able to test this important sector and thus shed light on the strong dynamics which generates the Higgs and other composite states. To elucidate our results we focus on the topological interactions of a minimal composite Higgs model with a fermionic ultraviolet completion, based on the coset $\SU(4)/\Sp(4)$. We suggest the strategy to test these interactions at the FCC-pp and analyse the expected reach.
\\[.3cm]
\end{abstract}

\maketitle

\section{Introduction}

Compelling theories which provide a natural explanation of the electroweak (EW) symmetry breaking involve a new strong gauge sector, from which the Higgs emerges as a composite pseudo Nambu-Goldstone-Boson (pNGB) state \cite{Kaplan:1983fs,Kaplan:1983sm,Dugan:1984hq}. Minimal realizations of this kind of extensions of the Standard Model with an underlying fundamental fermionic matter are based on the global flavor symmetry breaking pattern $\text{SU}(4) \to \text{Sp}(4) \sim\text{SO}(6) \to \text{SO}(5)$ \cite{Appelquist:1999dq,Duan:2000dy,Ryttov:2008xe,Katz:2005au,Gripaios:2009pe,Galloway:2010bp,Cacciapaglia:2014uja}.  A summary of the various possible theories of fundamental composite dynamics,  the link to first principle lattice results, and an in depth study of composite (Goldstone) Higgs dynamics stemming from the $\text{SU}(4) \to \text{Sp}(4) \sim\text{SO}(6) \to \text{SO}(5)$  pattern of chiral symmetry breaking can be found in \cite{Cacciapaglia:2014uja}. The associated gauged topological sector was first discussed in  \cite{Duan:2000dy,Lenaghan:2001sd} and summarised in \cite{Sannino:2009za} also in the context of near conformal dynamics.  

Besides the Higgs, the theory includes another composite pNGB, a CP-odd state $\eta$, and a pseudoscalar particle associated with the anomalous breaking of a global $\text{U}(1)$ symmetry in the new strong sector. In analogy with the corresponding state in QCD, this new particle will be called $\eta^{\prime}$. This class of theories is characterized by a topological structure which generates anomalous interactions among the composite scalars and the gauge bosons. They are described by the renowned gauged Wess-Zumino-Witten (WZW) action \cite{Wess:1971yu, Witten:1983tx} and depend on fundamental parameters of the underlying gauge dynamics. In particular, the specific structure of the interactions depend on the embedding of the EW sector in the coset and on the number of degrees of freedom characterizing the new strong sector. Thus, a study of the phenomenology associated with the topological interactions allows details of the underlying dynamics to be uncovered. This is similar to the case of the anomalous decay $\pi_0 \to \gamma\gamma$ which allowed for determining the number of colors in QCD. 
Previous studies \cite{Hill:2007nz, Gripaios:2007tk, Harvey:2007ca, Ferretti:2016upr, Tandean:1995ci} considered the phenomenology associated to topological interactions in theories of dynamical EW symmetry breaking.

In this work we study  the WZW interactions involving the $\eta$ and the $\eta^{\prime}$ states in the 
$\SU(4)/ \Sp(4)$ composite Higgs model and show that they play a major role in their production at the planned future circular proton-proton collider with a beam-colliding energy of 100 TeV, the FCC-pp \cite{Arkani-Hamed:2015vfh}. Here the relevant processes are given by the vector-boson-fusion (VBF) production mechanism in association with a $W/Z$ boson or an Higgs in the final state. We will describe a search strategy for detecting the process $\eta/\eta^{\prime}$ followed by a $W$ at the FCC-pp and present the collider reach on the fundamental parameters of the new strongly interacting theory. 

The paper is structured as follows. We  introduce the model  and the low-energy effective theory in section \ref{sec:model}. In section~\ref{sec:large-N} we discuss the limit with a large number of underlying new color degrees of freedom. In sections \ref{sec:wzw-terms} and \ref{sec:pheno} we report the anomalous WZW interactions and the phenomenology of the pseudoscalars $\eta$ and $\eta^{\prime}$. In section \ref{sec:analysis} we provide details of our analysis for the reach of the FCC-pp. Finally, we summarize our results in the concluding section \ref{sec:conclusion}.

\section{Underlying model}\label{sec:model}

We consider a composite Higgs model with an underlying fundamental  fermionic matter theory based on the coset $\text{SU}(N_f)/\text{Sp}(N_f)$.
Here $N_f$ represents the number of fundamental Weyl fermions, $\psi^a$, dubbed technifermions, which are taken in the pseudo-real representation of a new strong gauge group $G_\TC$. The most economical choice, at the level of the underlying theory, which allows for  a viable realization of the composite Higgs scenario, corresponds to $N_f=4$ and a symplectic gauge group 
\cite{Appelquist:1999dq,Duan:2000dy,Ryttov:2008xe,Katz:2005au,Gripaios:2009pe,Galloway:2010bp,Cacciapaglia:2014uja}. The new strong sector is described by the Lagrangian
\begin{equation}
	\mathcal{L}_{\TC} = - \dfrac{1}{4}\mathcal{G}^{A}_{\mu\nu}\mathcal{G}^{A,\mu\nu} + i\overline{\psi}_a \overline{\sigma}^\mu D_\mu \psi^{a} - \tfrac{1}{2} \left(\psi^{a} m_{ab} \epsilon_\TC  \psi^{b} \hc\right)\,,
	\label{eq:L_TC}
	\end{equation}
where  the gauge invariant mass terms $m_{ab}$ provides an explicit breaking of the global  $\SU(4)$ flavor symmetry.  In the limit of $m_{ab}\to 0$ the theory has an additional anomalous $\text{U}(1)$ global symmetry,  analogous to the $\text{U}(1)$ axial symmetry of  QCD.

$\SU(4)$ is dynamically broken to $\Sp(4)$ via the fermionic bilinear condensate
\begin{equation}
	\left\langle \psi^{a} \epsilon_{\TC} \psi^{b} \right\rangle = f^2 \Lambda \,\Sigma_0^{ab} \, ,
	\label{eq:fermion_condensate}
\end{equation}
where $\Lambda\approx 4\pi f$ is the typical composite scale of the theory and $f$ denotes the corresponding pion decay constant. Notice that $\Lambda$ in this context represents a simple estimate of the scale where the composite theory breaks down. This should not be taken as a rigorous cutoff, as a precise evaluation requires non-perturbative methods. In our study we will implicitly assume that the true cutoff of the theory is larger than the masses of the composite states. The unitary matrix $\Sigma_0$ is in a two-index, antisymmetric representation of  $\SU(4)$, and takes the general form~\cite{Galloway:2010bp}
\begin{equation}
	\Sigma_0 = \cos\theta\, \Sigma_B \,+\, \sin\theta \,\Sigma_H = \begin{pmatrix}	i\,\sigma_2 \cos\theta  & \mathbf{1}\sin\theta \\ - \mathbf{1}\sin\theta & -i\,\sigma_2 \cos\theta \end{pmatrix}\,,\label{eq:vacuum}
	\end{equation}   
where $\theta$  parametrizes the alignment between the two physically inequivalent vacua of the theory, $\Sigma_B$ and
$\Sigma_H$ \cite{Cacciapaglia:2014uja}. 
To achieve a realistic scenario for EW symmetry breaking we embed the fundamental technifermions in vectorial representations of 
 the SM gauge group $\SU(3)_\text{c}\times\SU(2)_\LL \times \U(1)_Y$, in order to  avoid gauge anomalies. In particular, we assign the four $\psi^a$ fields  to the representations  $(1,2)_0 \oplus (1,1)_{-1/2} \oplus (1,1)_{+1/2}$. From (\ref{eq:vacuum}) we have that the direction $\theta=0$ (the vacuum $\Sigma_B$) preserves the EW symmetry, while for $\theta=\pi/2$ (the vacuum $\Sigma_H$) the EW symmetry is fully broken (technicolor limit). The physical value of the vacuum alignment angle $\theta$ will depend on the effective potential of the
 theory below $\Lambda$, which is generated by the sources of explicit $\SU(4)$  breaking. In most of the composite Higgs scenarios, the global flavor symmetry is broken by EW gauge interactions and the couplings between the new strong sector and the SM fermions (in particular the top quark), in addition to the mass term of the fundamental technifermions, $m_{ab}$.

At energies below the condensation scale $ \Lambda$, the new degrees of freedom of the theory consist of composite
particles. The lightest resonances in this case are expected to be the pNGBs associated with the dynamical breaking of
$\SU(4)\to\Sp(4)$. In particular, there are five broken generators of the global flavor symmetry which are associated with the 
technipion fields $\pi_{1,2,3} $, the Higgs boson  $h$ and a new pseudoscalar $\eta$. After EW symmetry breaking, the three 
technipions become the longitudinal polarizations of the weak gauge bosons. We also include in the effective description the pseudoscalar composite state related to the anomalous $\text{U}(1)$ global symmetry, dubbed $\eta^\prime$, in analogy with the corresponding QCD meson. Thus, the low-energy theory is described by a non-linear $\Sigma$-model with
	\begin{equation}\label{Sigma}
	\Sigma = \exp\left[\dfrac{i}{f}\left(\tfrac{1}{2\sqrt{2}} \eta' + \pi_i Y^{i} + h Y^4 + \eta Y^5 \right)\right] \Sigma_0,
	\end{equation}
where $ Y^{i} $ are the broken generators of $\SU(4)$, cf.~Appendix~\ref{app:generators}.  Notice that in the expression (\ref{Sigma}) the $\eta^\prime$ 
is included in this way only in the limit of large $N$  \cite{tHooft:1973alw,Witten:1979kh}, $N$ being the number of degrees of freedom of the underlying gauge theory (see section \ref{sec:large-N}). In this case the  $\eta^\prime$ decay constant is $f_{\eta^\prime}=f(1+\mathcal{O}(1/N))$, and the mass of the $\eta^\prime$
is generated dominantly by non-perturbative instanton effects~\cite{Witten:1979vv,Veneziano:1979ec} (see also \cite{DiVecchia:2013swa}). In the large $ N $ limit it reads~\cite{DiVecchia:1980yfw,DiVecchia:1980xq}
	\begin{equation} \label{eq:etap-mass}
	m_{\eta^\prime}\approx \sqrt{\dfrac{2}{3}} \dfrac{f}{f_\pi} \dfrac{3}{N} m_{\eta_0} , 
	\end{equation}
where $ f_\pi = 92 \, \mathrm{MeV}$ is the pion decay constant and $ m_{\eta_0} = 849 \, \mathrm{MeV} $ is the mass of the QCD $ \SU(3) $-flavor singlet state $ \eta_0 $.	
It should be kept in mind  that \eqref{eq:etap-mass} is just an estimate of the instanton contribution to the $\eta^{\prime}$ mass.

 The gauged kinetic term of the effective Lagrangian is 
\begin{equation}
	\mathcal{L}_{\mathrm{kin}} = -\frac12 \Tr\left[(W^{i}_{\mu\nu})^\dagger W^{i\mu\nu} \right]-\frac 14 B_{\mu\nu} B^{\mu\nu}+f^2\Tr\left[\left(D_\mu \Sigma\right)^{\dagger} D^\mu \Sigma \right]\,
	\label{eq:L_kin}
	\end{equation} 
where $V_{\mu\nu}$ ($V=W^i,B$) denote the field strength of the EW gauge bosons and
\begin{equation}
	\quad D_\mu \Sigma =\partial_\mu \Sigma - i\mathcal{A}_\mu \Sigma -i\Sigma \mathcal{A}_\mu\transpose\,, 
\end{equation}
with
\begin{eqnarray}	
\label{EWgaugefields}
	\mathcal{A}_\mu & =& \dfrac{g}{2} W_\mu^{i} \begin{pmatrix} \sigma^{i} & \mathbf{0} \\ \mathbf{0} & \mathbf{0} \end{pmatrix} - \dfrac{g'}{2} B_\mu \begin{pmatrix} \mathbf{0} & \mathbf{0} \\ \mathbf{0} & \sigma^3 \end{pmatrix}\\
	& = &  \dfrac{g}{\sqrt{2}} \begin{pmatrix} W^{\prime+}_\mu \sigma^- + W^{\prime-}_\mu \sigma^+& \mathbf{0}\\ \mathbf{0} & \mathbf{0} \end{pmatrix} +\dfrac{g}{2}\left(s_w A^\prime_\mu+c_w Z^\prime_\mu\right)  \begin{pmatrix} \sigma^{3} & \mathbf{0} \\ \mathbf{0} & \mathbf{0} \end{pmatrix}+
	\dfrac{g'}{2} \left(s_w Z^\prime_\mu-c_w A^\prime_\mu\right) \begin{pmatrix} \mathbf{0} & \mathbf{0} \\ \mathbf{0} & \sigma^{3} \end{pmatrix}\,. \nonumber
\end{eqnarray}
In \eqref{EWgaugefields} we introduced the new EW gauge fields $W^{\prime\pm}$, $Z^\prime$ and $A^\prime$ 
in the basis rotated by the weak mixing angle
$\theta_w$, with  $s_w=\sin \theta_w$ and  $c_w=\cos\theta_w$.
We also defined $\sigma^\pm\equiv (\sigma^1 \mp i\, \sigma^2)$,  $\sigma^k$ ($k=1,2,3$) being the usual 
Pauli matrices. 

We expand the effective Lagrangian (\ref{eq:L_kin}) 
in the low momentum of the scalar fields, $p\ll \Lambda$,  up to $\mathcal{O}(p^2)$.
The bilinear terms involving the technipion fields $\pi_i$ in \eqref{Sigma}  read
\begin{eqnarray}
\begin{split}\label{Lkin2phys}
	\mathcal{L}_{\mathrm{kin}} = & -\frac12 W^{+}_{\mu\nu} W^{-\mu\nu} -\frac14 Z_{\mu\nu} Z^{\mu\nu} -\frac14 A_{\mu\nu} A^{\mu\nu} \\
	& + 2\,g^2\,f^2\,\sin^2\theta\, W^+_\mu\,W^{-\mu}  
	     + \frac{g^2}{c_w^2}\,f^2\,\sin^2\theta  \,Z_\mu \,Z^\mu+\ldots\,,
\end{split}
\end{eqnarray}
where we introduced the Stueckelberg fields
\begin{eqnarray}\label{eq:stuck2}
\begin{split}
	W^\pm_\mu & = W^{\prime\pm}_\mu\pm\frac{i}{\sqrt{2}\,g\,\sin\theta}\, \partial_\mu \pi^\pm \quad\quad \text{and} \quad\quad Z_\mu  = Z^\prime_\mu+\frac{c_w}{\sqrt{2}\,g\,\sin\theta}\,\partial_\mu \pi^3\,,
\end{split}
\end{eqnarray}
with the (massless) photon field $A_\mu\equiv A_\mu^\prime$ and $\pi^\pm\equiv (\pi^2\mp i\,\pi^1)/\sqrt{2}$. We interpret the fields $W^\pm$
and $Z$ as the physical massive weak gauge bosons with the corresponding masses given in the second line of 
\eqref{Lkin2phys}, provided one identifies the EW scale $v$ with
\begin{equation}
v \;=\; 2 \,\sqrt{2}\, f\, \sin \theta\;=\;246~\text{GeV}\,.
\end{equation}
 Notice that the terms shown in   \eqref{Lkin2phys} are equivalent to what one would obtain by performing a gauge transformation of the effective Lagrangian \eqref{eq:L_kin} where the technipion fields in $\Sigma$ are rotated away (\textit{unitary gauge}). Therefore,  the $\pi^i$ give the longitudinal polarizations of the EW gauge bosons.

In the unitary gauge  
the $\mathcal{O}(p^2)$ expansion of the effective Lagrangian \eqref{eq:L_kin} gives the additional interactions:
\begin{eqnarray}
	\begin{split}
	\mathcal{L}_{\mathrm{kin}}  \supset  &  \; \dfrac{1}{2}  (\partial_\mu h)^2 + \dfrac{1}{2}  (\partial_\mu \eta)^2 +  \dfrac{1}{2} (\partial_\mu \eta')^2\\ 
         + & \bigg( 2 g^2 W^{-}_\mu W^{+\mu} + (g^2 +g'^2) Z_\mu Z^\mu \bigg) \bigg(\tfrac{f}{2\sqrt{2}} \sin 2 \theta\,  h + \tfrac{1}{8} \cos 2\theta \,h^2 -\tfrac{1}{8} \sin^2\theta \, \eta^2 \bigg)\,.
	\end{split}\label{eq:L-kin-2}
	\end{eqnarray} 
Thus, the SM Higgs couplings to the gauge bosons are modified as
\begin{equation}
	g_{hWW} \equiv \dfrac{g^2}{\sqrt{2}}\,f\,\sin 2\theta \, = \, g^{\mathrm{SM}}_{hWW}\,\cos\theta \quad \andeq\quad g_{hZZ}\, \equiv \, g^{\mathrm{SM}}_{hZZ}\, \cos\theta\,.
	\end{equation}
Deviations from the SM couplings of the Higgs to EW gauge bosons are constrained by LHC Higgs data. The strongest bound is set by the combined ATLAS and CMS analysis and is obtained from the $ h\rightarrow Z Z $ decay channel \cite{Khachatryan:2016vau}, resulting in
	\begin{equation}
 			\sin\theta < 0.56 \quad \mbox{@ 95\% CL}\,,
	\end{equation}
which corresponds to
\begin{equation}\label{f-limit}
	f\; >\; 155~\text{GeV}\quad \mbox{@ 95\% CL}\,.
\end{equation}
The other couplings of the pNGBs to the gauge bosons from \eqref{eq:L-kin-2} read 
\begin{equation}
	g_{hhWW(ZZ)} \equiv   g^{\mathrm{SM}}_{hhWW(ZZ)}\,\cos 2\theta \andeq g_{\eta \eta WW (ZZ)} \equiv  \, g^{\mathrm{SM}}_{hhWW(ZZ)}\, \sin^2\theta\,.
\end{equation}

	Standard Model fermion masses can arise in this model by considering, e.g., 4-fermion interactions that could be generated, for example, via a  more fundamental chiral gauge theory dynamics \cite{Cacciapaglia:2015yra}. The simplest effective operator one can construct to give mass to the top is given by 
	\begin{equation}\label{eq:h-top}
	y_t' f \overline{Q}_\alpha \Tr\left[P^\alpha \Sigma \right] t_\RR \supset - \dfrac{m_{top}}{v}\,  \bar{t}_\LL \,t_\RR \left( v + \cos\theta\, h - i \sin\theta\, \eta'\right), 
	\end{equation}   
where $ \alpha $ is an $ \SU(2)_\LL $ index, and $ P^\alpha $ projects out the doublet of the pNGB matrix. We see that the operator also generates a tree-level coupling of the top to the Higgs and to the $\eta^{\prime}$, but it does not give a direct coupling to the $\eta$.\footnote{It does, however, generate the coupling $\eta\eta t \bar{t}$, which leads to $\eta$ pair production via top-mediated gluon fusion. The cross section for this process is small at the LHC with $\sqrt{s}$=14 TeV, but it becomes sizable at a 100 TeV collider, with a cross section of about 2 fb for $m_\eta=1$ TeV. We checked that this process does not represent a significant background for the analysis of topological interactions we will perform in section \ref{sec:analysis}.} Note that since we are considering technifermions which are singlets under the SM $\SU(3)_c$, our model does not include vector-like top-partners. The latter could serve to provide SM fermion masses through linear mass-mixing terms with elementary fermions, realizing the so-called partial compositeness scenario \cite{Kaplan:1991dc}.\footnote{Note that considering partial compositeness  in minimal composite Higgs model based on the coset $\text{SO}(5)/\text{SO}(4)$ and with top-partners in the spinorial representation of $\text{SO}(5)$, the so-called MCHM4 \cite{Agashe:2004rs}, the coupling of the Higgs to the top is the same as in eq. (\ref{eq:h-top}).} 
If, in addition to the modifications of the couplings to gauge bosons,  we consider the variation of the SM Higgs coupling to the top dictated by (\ref{eq:h-top}), the ATLAS analysis in \cite{Aad:2015pla} gives the bound

\begin{equation}\label{eq:h-stheta}
\sin(\theta)<0.35 \quad   \mbox{@ 95\% CL.}
\end{equation}

Non-zero mass terms for $h$ and $\eta$ are generated by the effective pNGB potential, which is induced by those interactions in the underlying theory that explicitly break the global $\text{SU}(4)$ flavor symmetry of the technifermions. The latter include EW gauge interactions and the effective couplings of the pNGBs to the SM fermions, in particular to the top quark. 
As a general feature of composite Higgs models \cite{Contino:2010rs, Bellazzini:2014yua}, the natural minimum of the potential occurs either at $\theta=0$ (driven by gauge contributions) or at $\theta=\pi/2$ (from the top interactions) \cite{Cacciapaglia:2014uja}. In order to achieve the EW symmetry breaking and a light 125 GeV Higgs, it is necessary to operate a fine-tuned cancellation among different terms in the potential which misalign the vacuum to $0 < \sin\theta \ll 1$. Thus, $\sin\theta$ gives an estimate of the size of fine-tuning in composite Higgs models. 
A hierarchy between the mass of the Higgs boson and the mass of the $\eta$, pushing the latter in the TeV range, can easily be achieved. For example, the $\eta$ can be directly coupled to SM fermions, in particular to the top \cite{Bellazzini:2015nxw}, and the fundamental technifermions can have a non-zero mass term, as in \eqref{eq:L_TC}. In 
this case, from the minimization of the effective potential one can have $m_\eta \approx m_h/\sin\theta$, as shown in \cite{Galloway:2010bp,Cacciapaglia:2014uja}.
Similarly, one could expect the mass of the $\eta^\prime$ to significantly deviate from the rough estimate given in \eqref{eq:etap-mass}. For this reason we will not limit our analysis to sub-TeV values of $m_{\eta/\eta^\prime}$. 

In the following we will assume that the minimum of the effective potential does not induce a vev for the $\eta$, thus preventing a mixing between the $\eta$ and the Higgs.~\footnote{The effect of this mixing is studied, e.g., in \cite{Sanz:2017tco}.} Similarly, we will assume that explicit breaking terms originated from higher dimensional operators in the effective theory will not generate a sizable mixing between the $\eta$ and the $\eta^\prime$.  
\mathversion{bold}
\section{The large-$N$ limit}\label{sec:large-N}
\mathversion{normal}
The potential for the pNGBs and the interactions among composite states are calculable under the assumption of a ``weakly"-coupled composite sector. This scenario could be realized by a large-$N$ strong dynamics, following the 't Hooft argument \cite{tHooft:1973alw}. In the case of $\SU(N)$ or Sp($N$) the mesons become asymptotically free in the limit $1/N \to 0$. In fact, at large $N$, the coupling $g_{\rho}$ of the mesonic interactions scales as 
\begin{equation}\label{eq:g-rho}
g_{\rho} \approx \frac{4\pi}{\sqrt{N}} \ .
\end{equation}
Similarly, we expect that the masses of composite spin-1 resonances follow the relation
\begin{equation} \label{eq:m-rho}
m_{\rho} \approx \dfrac{\sqrt{2}f}{f_\pi} \sqrt{\dfrac{3}{N}} m_{\rho_0},
\end{equation}
where $ m_{\rho_0} = 770 \, \mathrm{MeV} $ is the mass of the corresponding QCD vector meson. 
This estimate implies the following ratio between the mass of the spin-1 resonances and 
 the instanton-generated mass of the $\eta^{\prime}$ given in eq.~\eqref{eq:etap-mass}:
\begin{equation}
\frac{m_{\rho}}{m_{\eta^{\prime}}} \approx \sqrt{ N}  \ .
\end{equation}
At large $N$, we thus expect the spin-1 composite resonances to be considerably heavier than the $\eta^{\prime}$ (and of the $\eta$, considering that the latter is a pNGB). This will allow us to neglect their contribution in our subsequent analysis of topological terms.~\footnote{It should be clear that we keep fixed the new decay constant, and of course the EW scale as function of the number of new colors.  } 

Considering (\ref{eq:m-rho}), the $S$ parameter is naively given by  \cite{Galloway:2010bp, Arbey:2015exa} 
\begin{equation}
S\approx \sin^2\theta  \left( \frac{N}{4 \pi}- \frac{1}{6 \pi}\ln \frac{m_h}{\Lambda} \right)
\end{equation}
where the last term is the infrared contribution coming from the Higgs loop. A general and more detailed analysis  is performed in \cite{Sannino:2015yxa,Foadi:2012ga}. 
 The results of the global electroweak SM fit \cite{Baak:2014ora} gives $S= 0.05 \pm 0.11$,
which set an exclusion at 95\% CL on a region with large $N $ and $\sin\theta$ values. For example, taking $N=50\,  (30)$, values $\sin\theta \gtrsim 0.25 \, (0.33)$ are excluded. 
The custodial invariance of the flavor symmetry protects the model against large corrections to the $T$ parameter \cite{Sikivie:1980hm}. Electroweak constraints on the model are discussed in details in \cite{Galloway:2010bp,Arbey:2015exa}. 

It is important to point out that the 't Hooft argument applies to QCD-like dynamics and the extension of the scaling (\ref{eq:g-rho}) to a generic strongly-interacting theory must be considered only as a plausible guess. In our model, $N$ corresponds to the number of degrees of freedom associated to the underlying strong dynamics, which, in the case of one $ \SU(2)_\LL $ doublet of technifermions in the representation $R$ of $G_\TC$, is given by the dimension $d(R)$.

Theories of composite (Goldstone) Higgs at large $N$ are especially interesting because of the possibility of probing at colliders the underlying strong dynamics via their topological sector \cite{Appelquist:1999dq,Duan:2000dy,Hill:2007nz, Gripaios:2007tk, Harvey:2007ca, Sannino:2009za,Ferretti:2016upr}. Indeed, as we show in the following section, the WZW terms present a direct proportionality to $d(R)$.

\section{Topological sector of the theory}\label{sec:wzw-terms}

The gauged WZW action $\Gamma_{\WZW}$ can be expressed in terms of the differential forms \cite{Duan:2000dy,Sannino:2009za,Witten:1983tw} 
\begin{equation}
	\begin{split}
		&\mathcal{A} = \mathcal{A}_\mu \, \dd x^{\mu}, \quad \dd \mathcal{A} = \partial_\mu \mathcal{A}_\nu \,\dd x^{\mu} \dd x^{\nu},\\
		&\dd \Sigma = \partial_\mu \Sigma \, \dd x^\mu,\quad \alpha = \dd \Sigma \, \Sigma^{\dagger}.
		\label{eq:WZWnotation}
	\end{split}
\end{equation}
The full expression of $\Gamma_{\WZW}$ is given in Appendix \ref{app:WZW}.
We work in the unitary gauge where the pGBs $\pi_{i}$ are rotated away from the $\Sigma$ matrix and the gauge fields $\mathcal{A}$ are Stueckelberg fields, see eqs. \eqref{EWgaugefields} and \eqref{eq:stuck2}.

We start by considering the interactions between one scalar and two gauge bosons. These originate from the following terms of the WZW action 
\begin{equation}
	\Gamma_{\WZW}\supset -10 c \int \Tr\left[(\dd \mathcal{A} \, \mathcal{A} + \mathcal{A} \, \dd \mathcal{A})\alpha \right] -5 c \int \Tr \left[\dd \mathcal{A}\, \dd \Sigma\, \mathcal{A}\transpose \Sigma^\dagger - \dd \mathcal{A}\transpose\, \dd \Sigma^\dagger \, \mathcal{A} \Sigma \right],
	\label{eq:WZW-AdA}
\end{equation}    
with $ c= -i d(R)/480\pi^2 $. There are no anomalous couplings of the Higgs to two gauge bosons arising from these terms. Conversely, for the $\eta$ and the $\eta^{\prime}$ we get the following interactions: 
\begin{equation}
	 -\dfrac{d(R) \alpha_\EM \cos\theta \sin\theta}{32\pi v} \eta \left[\dfrac{2}{c_w s_w} A_{\mu\nu}\tilde{Z}^{\mu\nu} +\dfrac{c_w^2 -s_w^2 }{c_w^2 s_w^2} Z_{\mu\nu} \tilde{Z}^{\mu\nu} +\dfrac{2}{s_w^2} W^+_{\mu\nu} \tilde{W}^{-\mu\nu} \right]
	\label{eq:WZW-AAeta}
\end{equation}
and 
\begin{eqnarray}
	 & -\dfrac{d(R) \alpha_\EM \sin\theta}{48\pi v} \eta' \left[3 A_{\mu\nu}\tilde{A}^{\mu\nu} + 3\dfrac{c_w^2-s_w^2}{c_w s_w} A_{\mu\nu}\tilde{Z}^{\mu\nu} +\dfrac{3- 6 c_w^2 s_w^2 -\sin^2\theta}{2 c_w^ 2s_w^2} Z_{\mu\nu} \tilde{Z}^{\mu\nu}  \right. \nonumber \\ 
	& \left.+\dfrac{3- \sin^2\theta}{s_w^2} W^+_{\mu\nu}\tilde{W}^{-\mu\nu} \right].
	\label{eq:WZW-AAetap}
\end{eqnarray} 
Here the tensor fields are defined by $ V_{\mu\nu} = \partial_\mu V_\nu - \partial_\nu V_\mu $ and $ \tilde{V}^{\mu\nu} = \varepsilon^{\mu\nu\rho\sigma} V_{\rho\sigma} $, with $V=W^{\pm},Z,A$. Note that differently from the effective theories described in \cite{Gripaios:2009pe, Bellazzini:2015nxw} we do not have an electromagnetic anomaly for the $\eta$. 

The pseudoscalar interactions with three gauge bosons originate from the following terms: \footnote{The last term does not contribute in the case of the coset $\SU(4)/\Sp(4)$.}
\begin{equation}
	\Gamma_{\WZW} \supset 10i c  \int \Tr\left[\mathcal{A}^3\alpha \right] -10 i c\int \Tr \left[ \mathcal{A} \alpha \mathcal{A} \Sigma \mathcal{A}\transpose \Sigma^\dagger \right]  +10 i c\int \Tr\left[(\dd \mathcal{A}\, \mathcal{A} + \mathcal{A} \dd \mathcal{A})\Sigma \mathcal{A}\transpose \Sigma^{\dagger} \right],
	\label{eq:WZW-AAA}
\end{equation}    
and read
\begin{equation}
	\begin{split}
	&	-i\dfrac{d(R) \alpha^{3/2}_\EM \sin\theta }{12 \sqrt{\pi}v} \varepsilon^{\mu\nu\rho\sigma} \partial_\mu\eta' \left[\dfrac{6-2\sin^2\theta}{s_w^2}A_\nu + \dfrac{6 c_w^2 -(1+2c_w^2) \sin^2\theta}{c_w s_w^3} Z_\nu\right] W_\rho^+ W_\sigma^-\\
	& -i\dfrac{d(R) \alpha^{3/2}_\EM\cos\theta \sin\theta}{4\sqrt{\pi} v} \varepsilon^{\mu\nu\rho\sigma} \partial_\mu\eta \left[\dfrac{2}{s_w^2}A_\nu + \dfrac{2c_w^2 - \sin^2\theta}{c_w s_w^3}Z_\nu \right] W_\rho^+ W_\sigma^- .
	\end{split}
	\label{eq:WZW-AAAeta}
\end{equation}
Also in this case, no anomalous couplings for the Higgs are generated.
Finally, we derive couplings between two scalars and two gauge bosons, which arise from\begin{equation}
	\Gamma_{\WZW} \supset 5 c \int \Tr\left[(\mathcal{A} \alpha)^{2} \right] -5 c \int \Tr \left[ \Sigma \mathcal{A}\transpose \Sigma^{\dagger}(\mathcal{A}\alpha^2 + \alpha^2 \mathcal{A}) \right].
	\label{eq:WZW-AA}
\end{equation}   
In addition there will be contributions from the expansion of \eqref{eq:WZW-AdA}. We find 
	\begin{equation}\label{eq:h}
	\begin{split}
	 & -\dfrac{d(R)\alpha_\EM \cos\theta \sin^3 \theta }{24 \pi v^2} \varepsilon^{\mu\nu\rho\sigma} h \partial_\mu \eta'\left[\dfrac{1}{c_w^2 s_w^2}Z_{\nu\rho}Z_\sigma + \dfrac{1}{s_w^2} \left(W_{\nu\rho}^+ W_\sigma^- + W_{\nu\rho}^- W_\sigma^+ \right) \right]\\
	&- \dfrac{d(R)\alpha_\EM\sin^3 \theta }{16 \pi v^2} \varepsilon^{\mu\nu\rho\sigma} \bigg[\dfrac{4}{c_w s_w} \partial_\mu h \partial_\nu \eta A_\rho Z_\sigma \\
	&+ h\overleftrightarrow{\partial_\mu} \eta
	\left( \dfrac{c_w^2-s_w^2}{c_w^2 s_w^2}Z_{\nu\rho}Z_\sigma +\dfrac{1}{s_w^2} \left(W_{\nu\rho}^+ W_\sigma^- + W_{\nu\rho}^- W_\sigma^+ \right) +\dfrac{1}{c_ws_w} \left(A_{\nu\rho}Z_\sigma + Z_{\nu\rho}A_\sigma \right) \right)\bigg].
	\end{split}
	\end{equation}

It should be noted that the WZW action does not give rise to anomalous trilinear and quartic gauge couplings. This is due to the fact that the minimal scenario, with the fermion content defined in section \ref{sec:model}, is free from gauge anomalies. In the case of a different hypercharge assignment of the fundamental technifermions, EW gauge anomalies appear and must be cancelled by adding new fermions which are singlets under $G_{\TC}$ and have the appropriate hypercharge. In this case trilinear and quartic anomalous gauge couplings stemming from $\Gamma_\WZW$ are suppressed by the interference with the loop contributions of the new fermions.

\begin{figure}[t!]
\begin{center}
\includegraphics[width=0.45\textwidth]{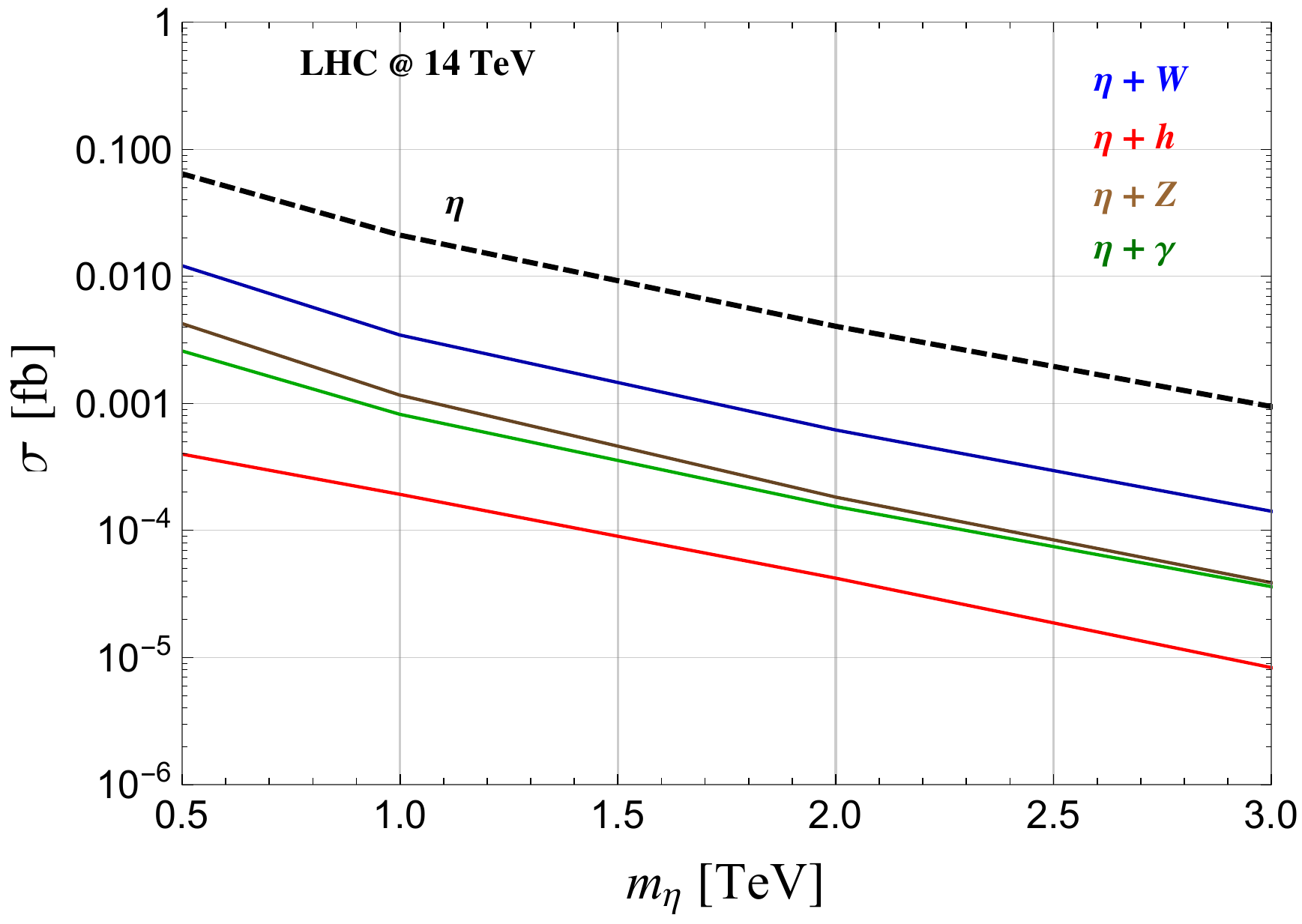}\quad \includegraphics[width=0.45\textwidth]{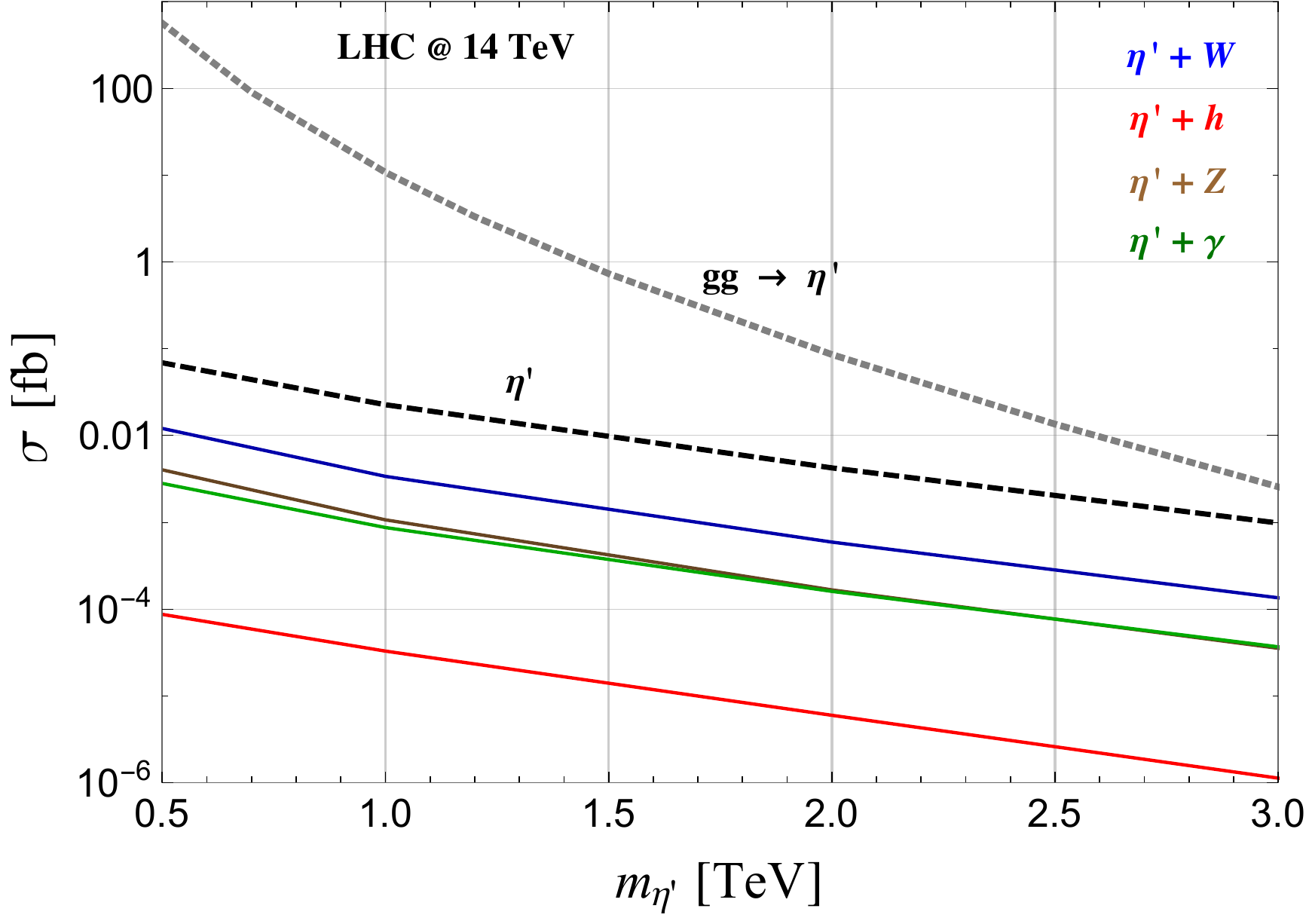} \\ \vspace{0.15cm}
\includegraphics[width=0.45\textwidth]{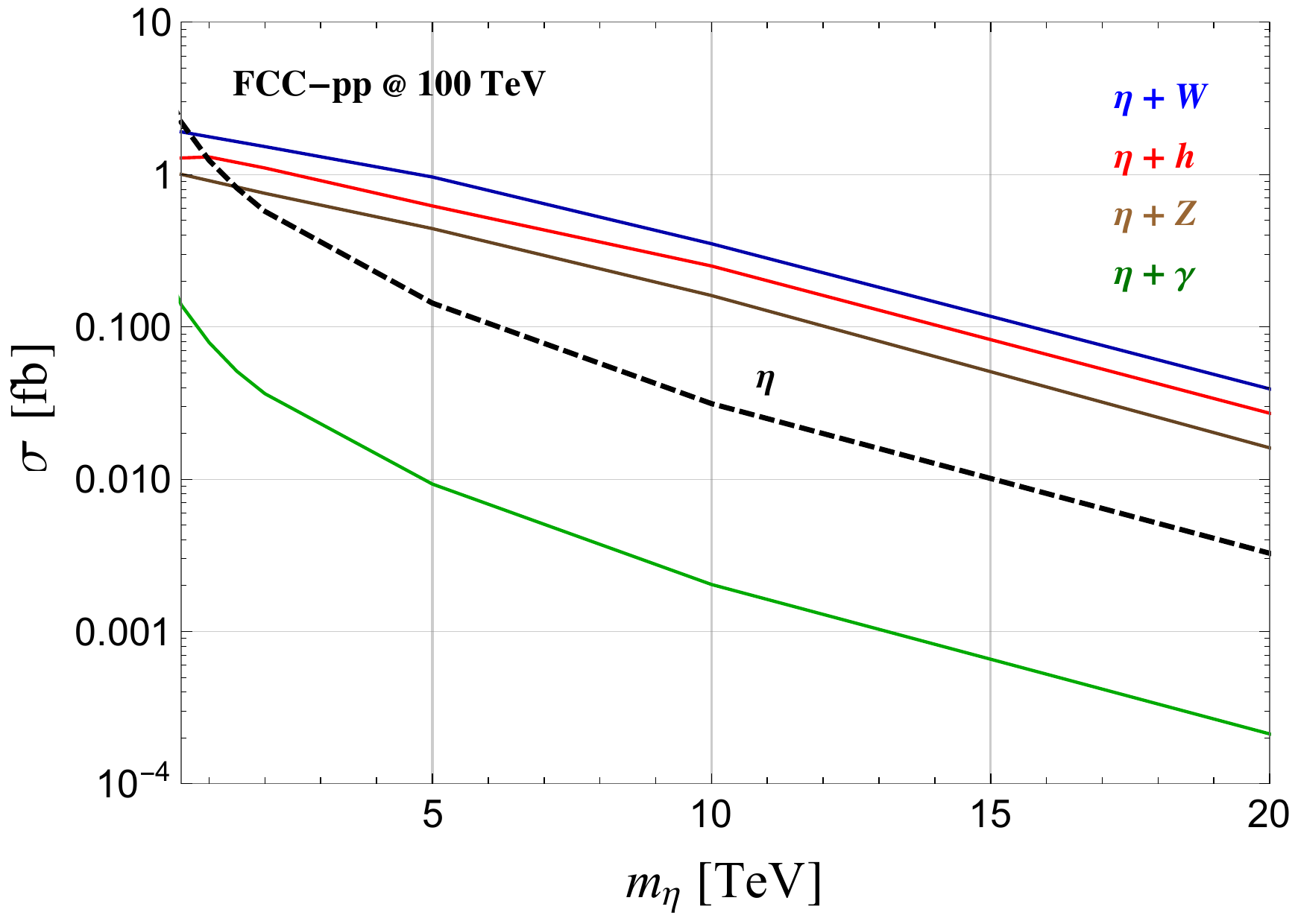} \quad
\includegraphics[width=0.45\textwidth]{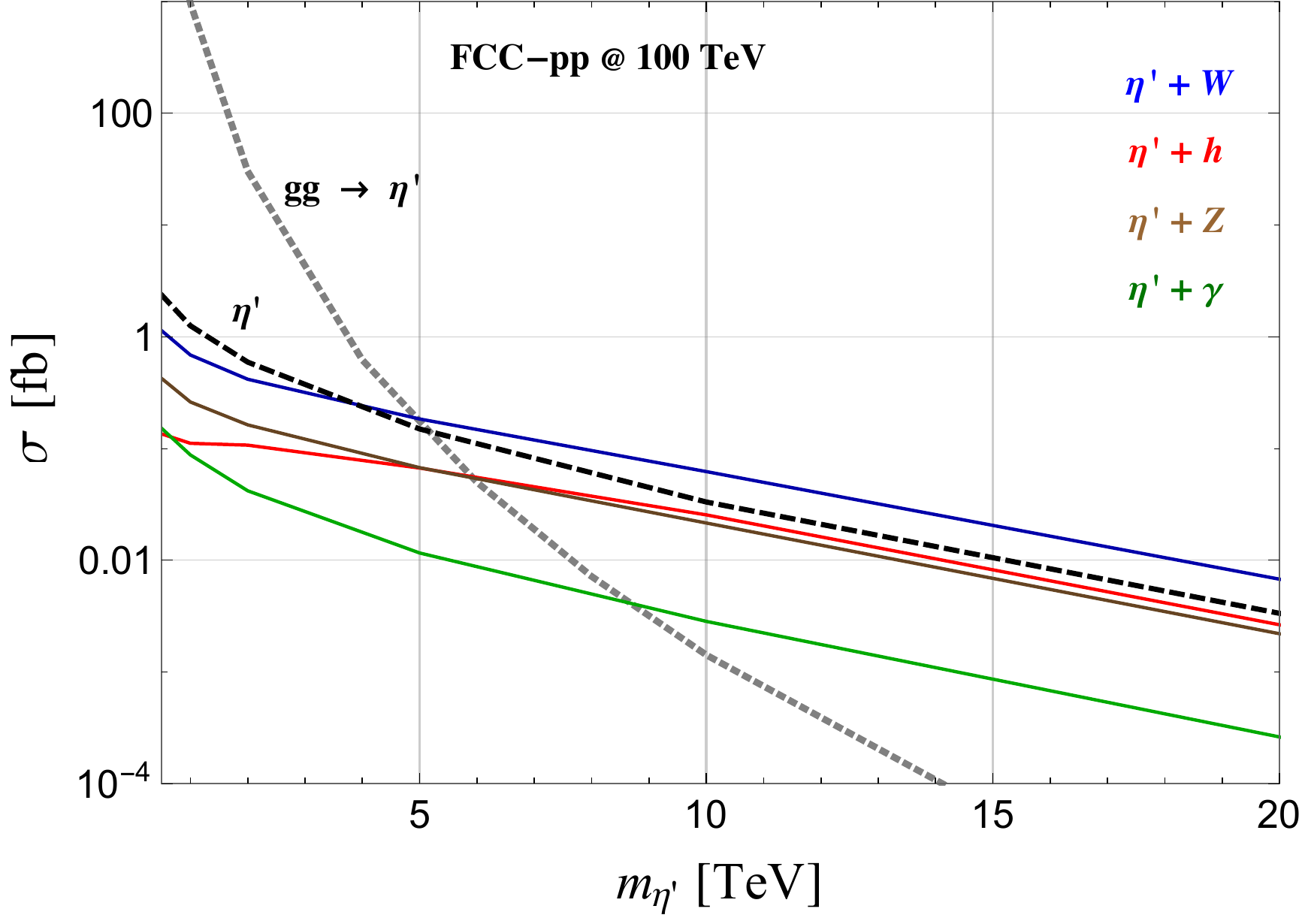}
\end{center}
\caption{ \label{fig:xsec}
Cross sections for different $\eta/\eta^{\prime}$ production mechanisms at the 14 TeV LHC ({\it upper panels}) and at the 100 TeV FCC-pp ({\it lower panels}). The black-dashed curve shows the VBF production resulting only from the pseudoscalar topological interaction with two gauge bosons (eqs.~\eqref{eq:WZW-AAeta} and \eqref{eq:WZW-AAetap}). The cross section for the $\eta/\eta^{\prime}$ production accompanied by the Higgs or by a gauge boson derives from the interference between processes mediated by topological interactions of the pseudoscalar with two (eqs.~\eqref{eq:WZW-AAeta} and \eqref{eq:WZW-AAetap}) and three particles (eqs.~\eqref{eq:WZW-AAAeta} and \eqref{eq:h}). We fix $d(R)=10$ and $\sin\theta =$ 0.25. The gray-dotted curves on the right panels indicate the production cross section for the $\eta^{\prime}$ from top-mediated gluon fusion. It assumes an $\eta^{\prime}$ coupling to the tops: $\sin\theta m_t/v$, as resulting from a 4-fermion operator for the top mass generation, see eq. \eqref{eq:h-top}. }
\end{figure}

The effects of the anomalous interactions of the scalar resonances with the SM gauge bosons are expected to be enhanced at high energy, eventually leading to a breakdown of perturbative unitarity. 
The largest constraints are obtained from the dimension-5 operators in eqs.~(\ref{eq:WZW-AAeta}) and (\ref{eq:WZW-AAetap}), which  contribute to  the $VV\to V^\prime V^\prime$
($V,V^\prime=W^\pm,Z,A$) scattering amplitudes.
From these processes  one can derive an upper bound on the scale where the effective description encoded in the perturbative expansion of the WZW action needs to be ultraviolet completed. Indeed,  at energies $\sqrt{\hat s}\gg m_{\eta/\eta^\prime}$, the dominant contribution to the scattering amplitude of gauge bosons 
is given by the s-channel exchange of the scalar resonances, which in the case of the $\eta$ leads to the constraint
\begin{eqnarray}
	\sqrt{\hat s} & \lesssim & m_\eta\, \Bigg( \frac{\Gamma(\eta\to W^+W^- )}{m_\eta}\,+\,\frac{\Gamma(\eta\to ZZ)}{m_\eta} \Bigg)^{-1/2} \; \approx \;\; \frac{800~\text{TeV}}{d(R)\,\sin\theta}\,,\label{bound-unitarity}
\end{eqnarray}
with the expressions for the partial decay rates of the $\eta$ given in Appendix~\ref{app:rates}.
A similar bound is obtained from the exchange of the $\eta^\prime$, for which there is an additional contribution coming from the diphoton decay rate. We remark that the right-hand-side of (\ref{bound-unitarity})
is  a conservative estimate of the cutoff of the effective theory. 
In theories of strong gauge dynamics there is a physical scale, the composite scale $\Lambda$, 
above  which the effective theory is not well defined. As pointed out earlier, the actual value of $\Lambda$ can only be computed using non-perturbative methods.


\mathversion{bold}
\section{Production mechanisms for the $\eta$ and the $\eta^{\prime}$ }\label{sec:pheno}
\mathversion{normal}

We now discuss aspects of the $\eta/\eta^{\prime}$ phenomenology which will be relevant to our subsequent analysis.
In particular, we will concentrate on processes involving $\eta/\eta^{\prime}$ anomalous interactions as a way to uncover topological terms at future colliders. We refer the reader to the studies in refs. \cite{Gripaios:2009pe, Galloway:2010bp, Molinaro:2015cwg, Molinaro:2016oix, Bellazzini:2015nxw} for details on the phenomenology of the $\eta$ and $\eta^{\prime}$ states.

If the $\eta/\eta^{\prime}$ does not couple directly to fermions, the dominant production mechanism is vector-boson-fusion (VBF) mediated by topological interactions. The $\eta$ can also be doubly-produced via VBF through the non-topological terms in $\mathcal{L}_{\rm kin}$, eq. (\ref{eq:L_kin}). We find that the cross sections for the single and double production of these composite states at the LHC are small (see Fig. \ref{fig:xsec}) and no significant constraints are currently placed by the LHC \cite{Molinaro:2016oix, Khachatryan:2016yec, ATLAS:2016eeo, ATLAS:2016npe, CMS:2017fge,CMS:2017wsr}. 
If tree-level interactions with the top are present, the $\eta/\eta^{\prime}$ is dominantly produced at the LHC (and at the FCC-pp, for masses typically lighter than 5 TeV) by top-mediated gluon fusion, as shown in Fig.~\ref{fig:xsec}, 
and fully decays into $t \bar{t}$ pairs.\footnote{The gluon fusion cross section is calculated at next-to-leading order in QCD, using the implementation given in \cite{Demartin:2014fia}.}~Even in this case, LHC does not yet provide significant constraints \cite{Molinaro:2016oix, Sirunyan:2017uhk, TheATLAScollaboration:2016wfb}.  
Figure \ref{fig:xsec} shows the cross sections for different $\eta/\eta^{\prime}$ production mechanisms at the 14 TeV LHC (upper panels) and at the 100 TeV FCC-pp (lower panels), as a function of the pseudoscalar mass. We apply minimal cuts on the transverse momenta and pseudorapidities of the two final jets: $p_T (j) >$ 20 GeV and $|\eta_j|<$5 (6) for the LHC (FCC-pp). We fix $d(R)=$10 and $\sin\theta=$0.25. The black-dashed curve shows the VBF production of the $\eta/ \eta^{\prime}$ resulting only from the pseudoscalar topological interactions with two gauge bosons (see eqs.~(\ref{eq:WZW-AAeta}) and (\ref{eq:WZW-AAetap})). The colored continuous curves indicate the $\eta/\eta^{\prime}$ production accompanied by the Higgs or by a gauge boson. These processes are mediated by topological interactions with four bosons (eqs.~(\ref{eq:WZW-AAAeta}) and (\ref{eq:h})), which interfere with the terms in the WZW action involving only three bosons (see eqs. (\ref{eq:WZW-AAeta}) and (\ref{eq:WZW-AAetap})), as shown in Fig. \ref{fig:S-diagram}.
Our calculations show that the production cross section of pseudoscalars accompanied by an EW gauge boson or a Higgs at the 100 TeV collider are highly enhanced compared to the LHC and are even larger than the cross section for the $\eta/\eta^{\prime}$ single production resulting from \eqref{eq:WZW-AAeta} and \eqref{eq:WZW-AAetap}.\footnote{Notice that the cross section for $\eta/\eta^{\prime}+\gamma$ is smaller than $\eta/\eta^{\prime}+W,Z$. This is in part due to the cuts applied to the photon, $p_T \, (\gamma)>$ 10 GeV, $|\eta_\gamma|<2.5$, and to the suppression $\tan^2\theta_w$ in the $\eta/\eta^{\prime} \gamma  WW$ coupling compared to $\eta/\eta^{\prime}Z WW$.} This enhancement is due to the derivative coupling structure of the vertices entering in diagrams similar to those in Fig. \ref{fig:S-diagram}, which give the production of the pseudoscalars associated with Higgs and gauge bosons. In our study we will perform a detailed analysis of the $\eta$ and $\eta^{\prime}$ production followed by a $W$ at the 100 TeV FCC-pp, which will allow us to indicate the reach on the main parameters characterizing the underlying strong dynamics, $d(R)$ and $\sin\theta$.

\section{FCC-pp reach on topological interactions}\label{sec:analysis}

\subsection{The signal}
\begin{figure}
\begin{center}
\begin{minipage}[t]{.48\linewidth}
\includegraphics[width=0.7\textwidth]{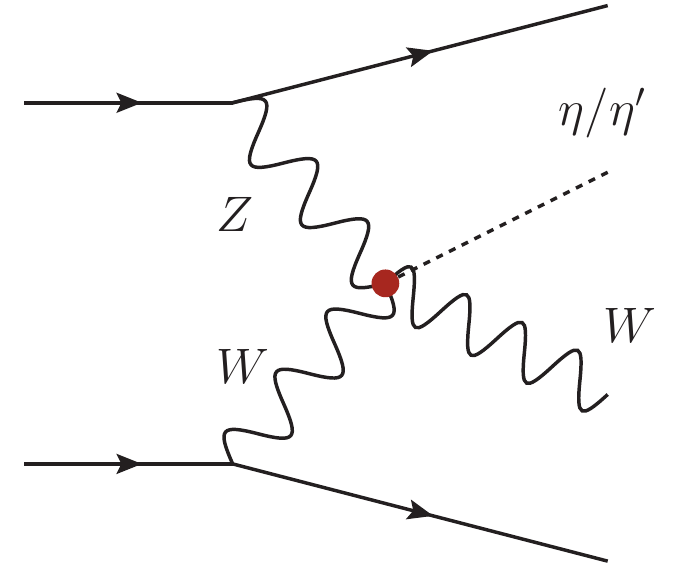}
{ \footnotesize \\Diagram containing a 4-boson topological vertex (eq.~(\ref{eq:WZW-AAAeta})). }
\end{minipage}
\begin{minipage}[t]{.48\linewidth}
\includegraphics[width=0.7\textwidth]{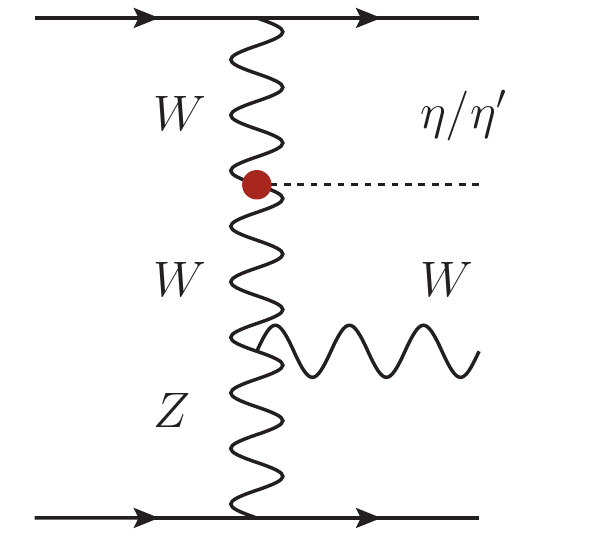}
{ \footnotesize  \\Diagram containing a 3-boson topological vertex (eqs.~(\ref{eq:WZW-AAeta}) and (\ref{eq:WZW-AAetap})).}
\end{minipage}
\caption{ \small Relevant diagrams for the production via VBF of an $\eta/\eta^{\prime}$ particle accompanied by a $W$ boson.}
\label{fig:S-diagram}
\end{center}
\end{figure}

We focus on the production via VBF of an $\eta$ or an $\eta^{\prime}$ particle accompanied by a $W$ boson. This process is mediated by topological interactions. Specifically, it results from the interference between diagrams containing topological couplings with three and four bosons, as indicated by the red dot in Fig. \ref{fig:S-diagram}.

%
%
We consider the case where the $\eta$ does not directly couple to SM fermions while the $\eta^{\prime}$ interacts with the top quark, as resulting from the scenario where the top mass is generated from a 4-fermion operator, see eq.~(\ref{eq:h-top}). The $\eta^{\prime}$ thus decays dominantly to $t\bar{t}$, while the $\eta$ decays to $W^{+}W^{-}$, $ZZ$ and $Z\gamma$ through the topological interactions in (\ref{eq:WZW-AAeta}), as shown in the Appendix~\ref{app:rates}. In the case of very massive pseudoscalars, $m_{\eta/\eta^{\prime}}\gtrsim 5$ TeV, the 3-body decays into gauge bosons, mediated by 3- and 4-boson topological interactions, become relevant (see, e.g., \cite{Molinaro:2016oix}). We find that the 3-body decay rate represents about 50\% of the total decay width of a 5 TeV $\eta$. For an $\eta^{\prime}$ of 5 (10) TeV, the branching ratio for the decay into three gauge bosons is 4 (37)\%, in the case where the $\eta^{\prime}$ interacts with the top as in \eqref{eq:h-top} and  $d(R)=10$. Our study is focused on 2-body decays and pseudoscalars up to 5 TeV. For heavier $\eta/\eta^{\prime}$, it is necessary to tailor the analysis for the dominant 3-body decay.  We leave this to a future study. 

We consider the subsequent fully-hadronic decay of the $\eta$ and the $\eta^{\prime}$ for masses in the range 500 GeV $\leq m_{\eta/\eta^{\prime}} \leq$ 5 TeV. We apply a jet reconstruction procedure which identifies a (mostly) single fat-jet coming from the pseudoscalars in the case where they directly decay either to gauge bosons (via topological terms) or to tops. The signal acceptance is very similar in these two scenarios. We further consider the leptonic decay of the $W$. 
 The final state is given by
\begin{equation} \label{eq:final-state}
\ell+ n \, \text{jets} \, + \slashed{E}_T \, \, , \qquad n \geq 3 \, ,\qquad \ell \equiv e,\mu \ ,
\end{equation}
where $\slashed{E}_T$ indicates the transverse missing energy.
We apply the following acceptance cuts:
\begin{equation} \label{eq:acceptanceCuts}
\begin{split}
& p_T(\ell)  > 300 \, \text{GeV} \, , \quad \slashed{E}_T > 300 \, \text{GeV} \, , \quad p_T(j) > 20 \, \text{GeV} \, , \\
& |\eta_j|<6 \, ,\quad |\eta_\ell|< 2.5 \, ,\quad \Delta R (j-j)>0.4\, , \quad \Delta R (j-\ell)>0.4\, ,
\end{split}
\end{equation}
with $\Delta R (a-b)$ denoting the standard angular separation between particles $a$ and $b$.
We choose a hard acceptance cut on the lepton transverse momentum and on missing energy to significantly reduce the background, while retaining the majority of the signal events (the acceptance to the cuts is more than 80\% for the signal). We consider a pseudorapidity acceptance for the jets slightly larger than the one at the LHC ($|\eta_j|<6$ compared to 5) since, as also highlighted in recent studies for the FCC-pp \cite{Golling:2016gvc, Mohan:2015doa, Goncalves:2017gzy}, it significantly increases the reach on VBF processes. 

\begin{figure}[t!]
\begin{center}
\includegraphics[width=0.45\textwidth, height=5.2cm]{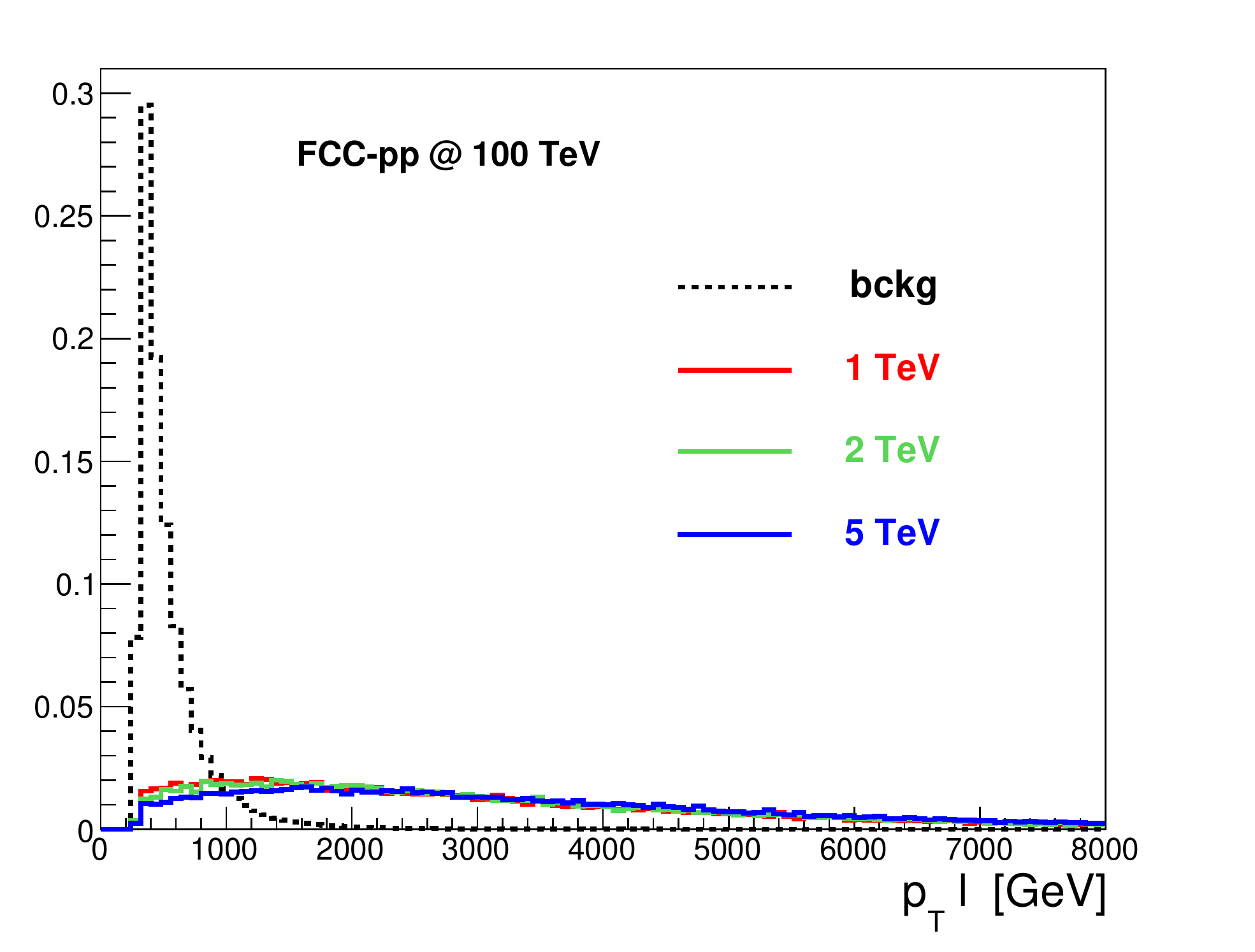}
\includegraphics[width=0.45\textwidth, height=5.2cm]{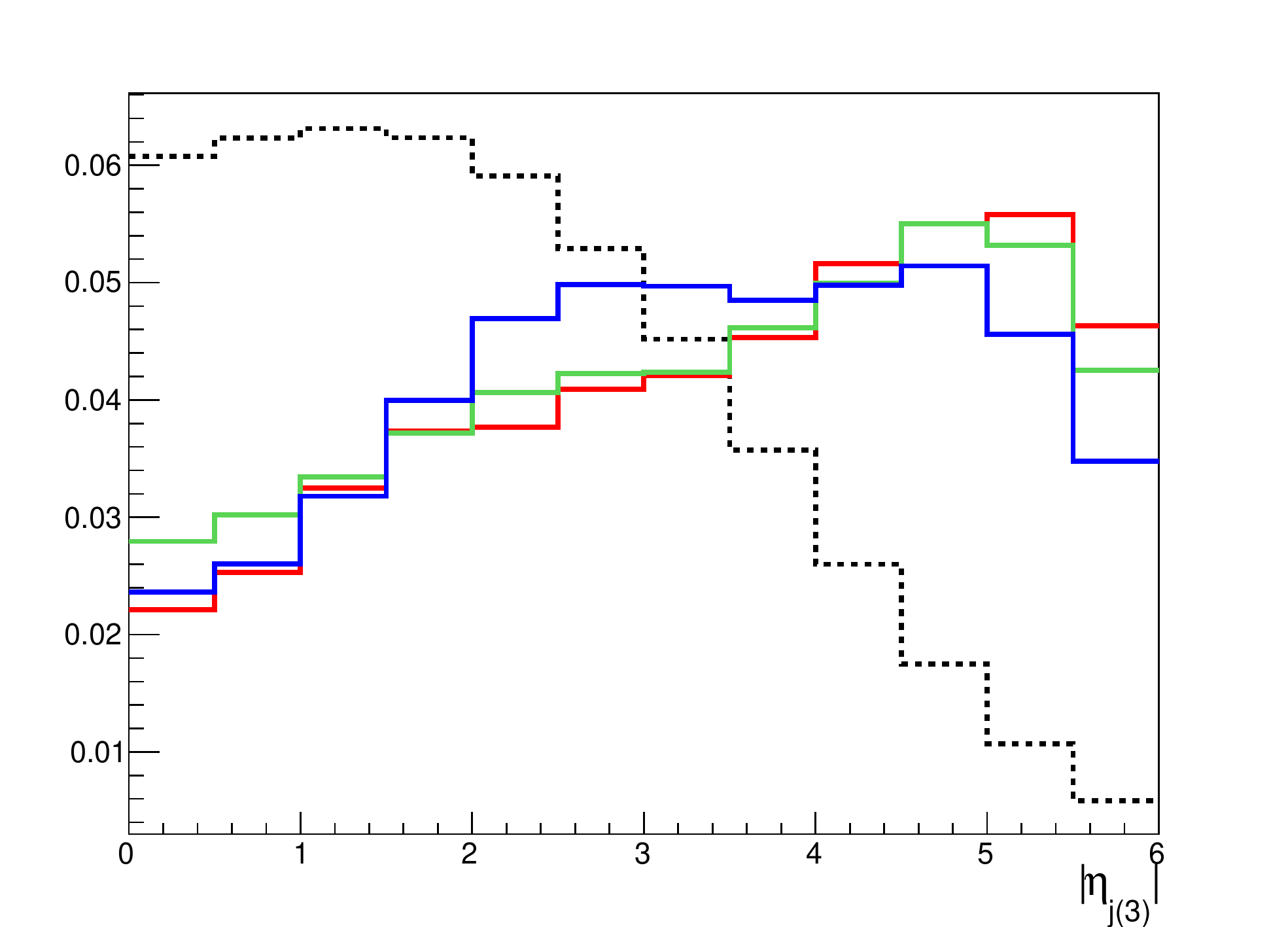}\\
\includegraphics[width=0.45\textwidth, height=5.2cm]{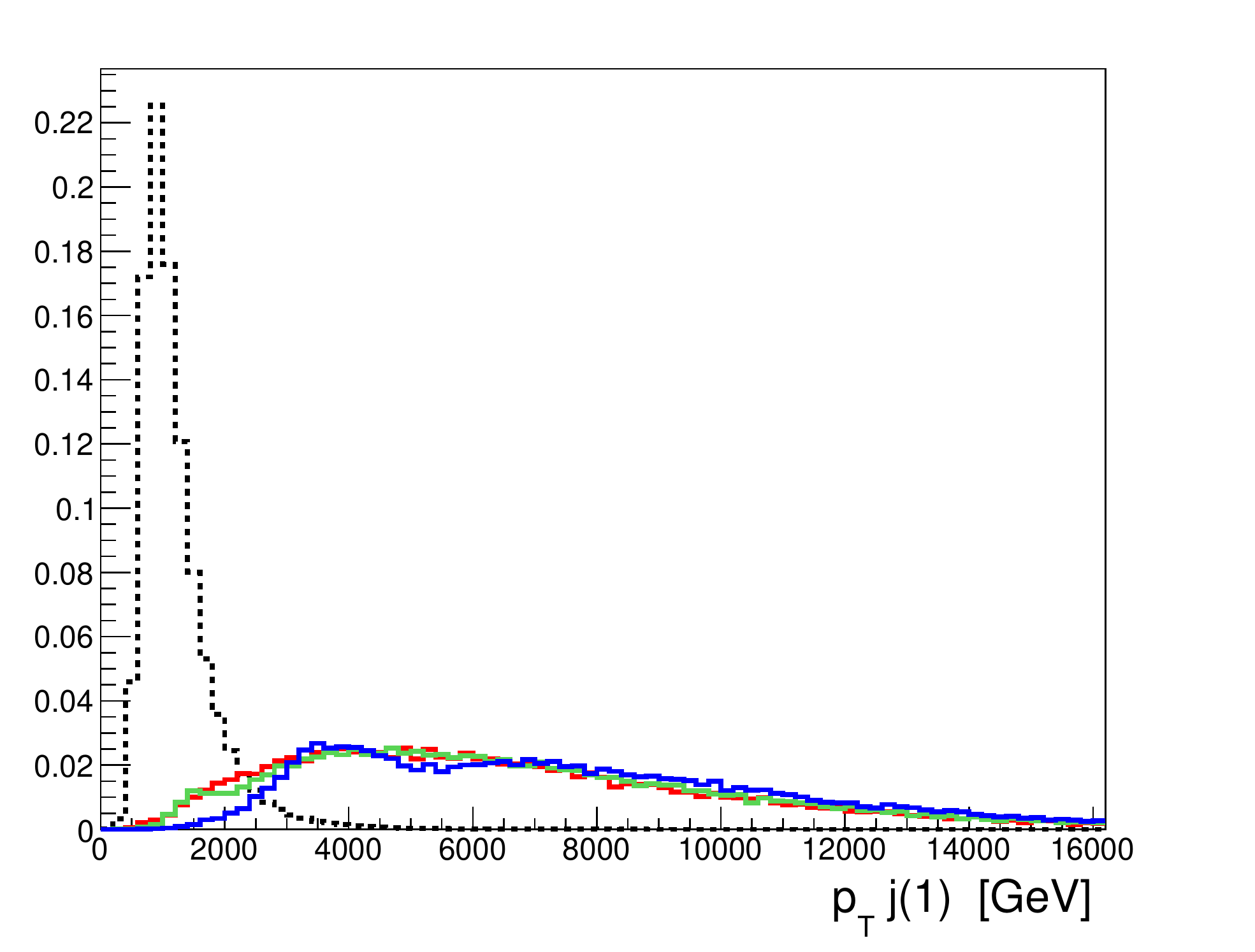}
\includegraphics[width=0.45\textwidth, height=5.2cm]{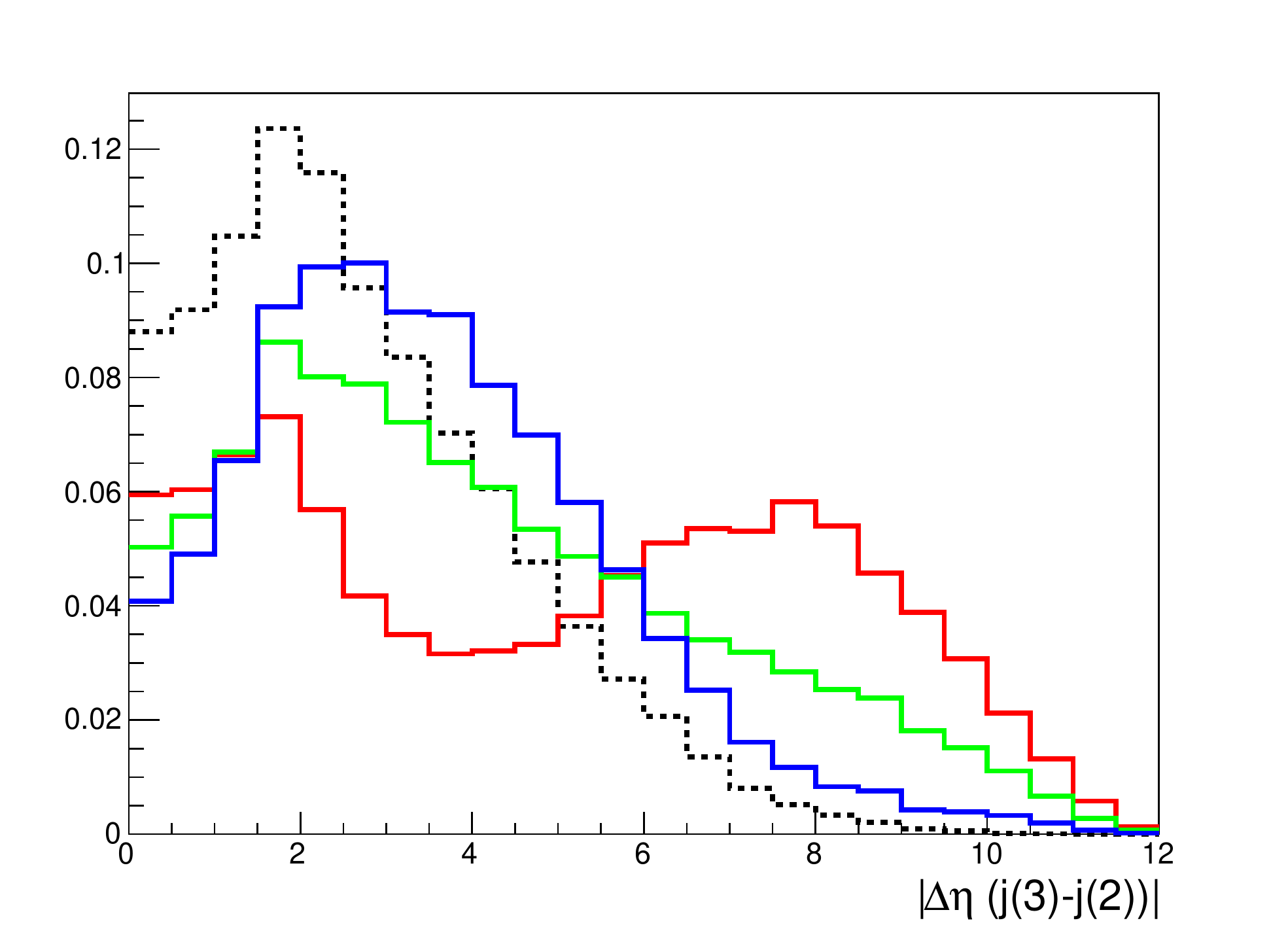}\\
\includegraphics[width=0.45\textwidth, height=5.2cm]{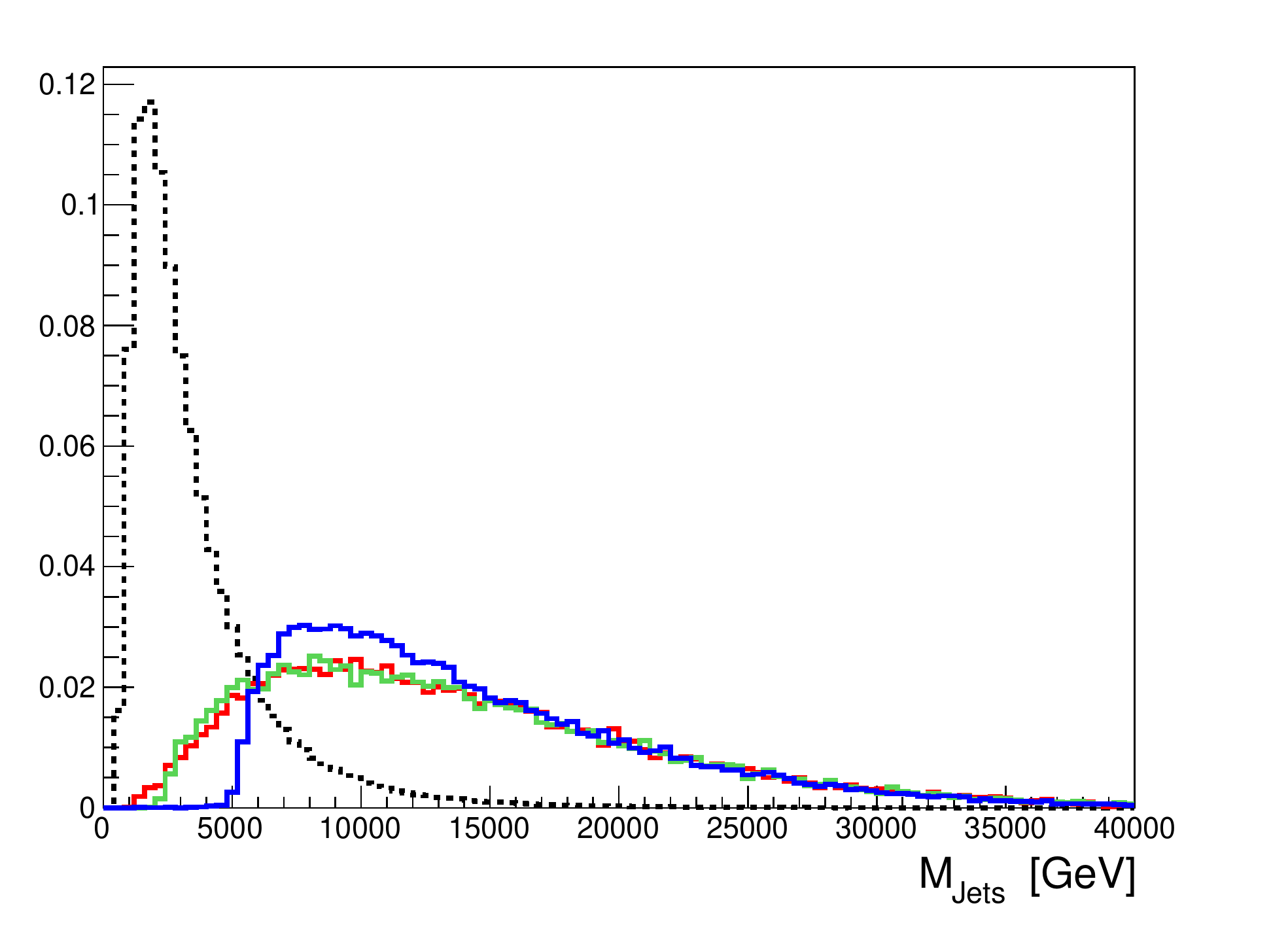}
\includegraphics[width=0.45\textwidth, height=5.2cm]{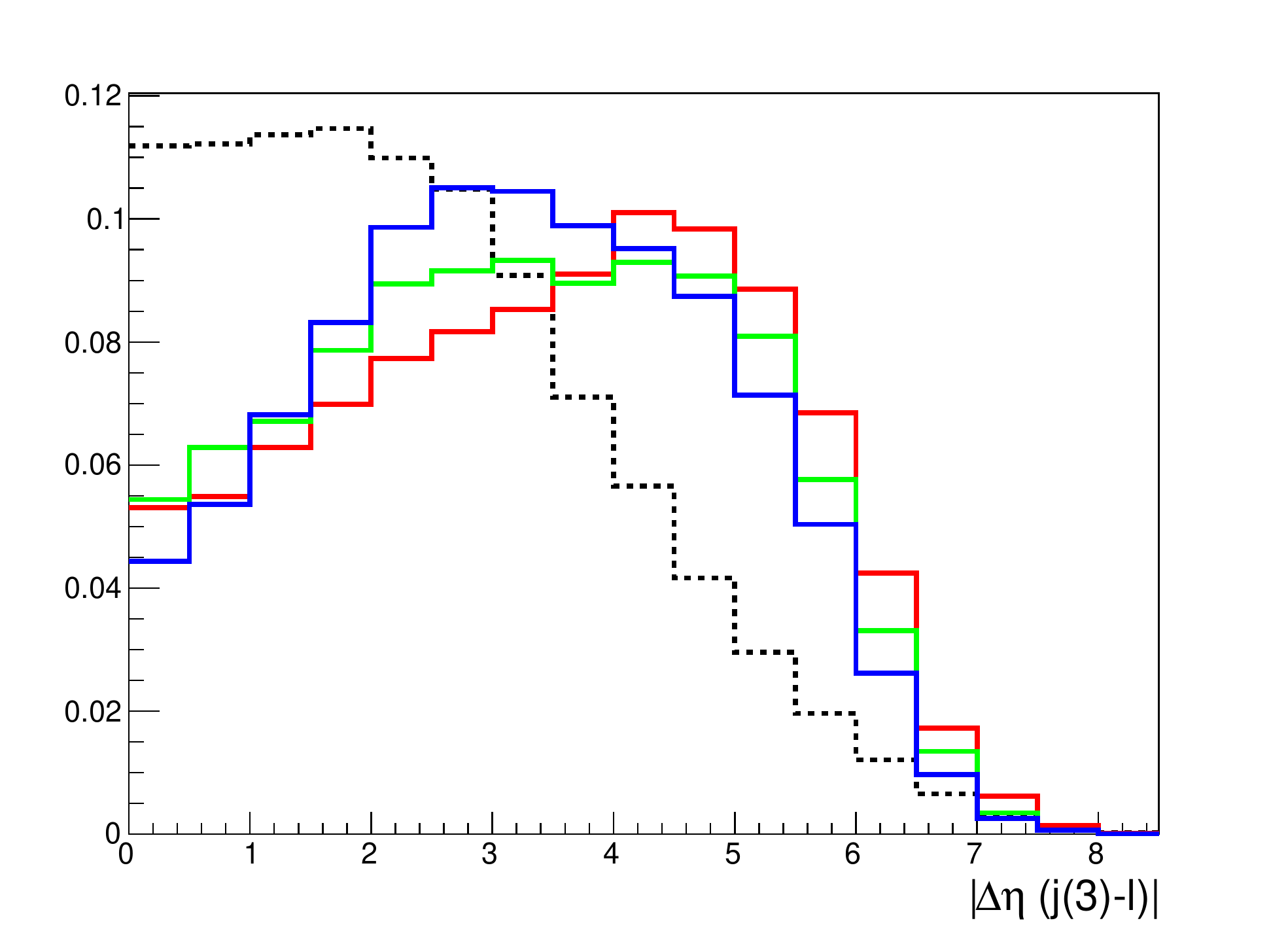}
\caption{ \small Distributions normalized to unit area for background (dotted-black curve) and signal with different $\eta$ mass values of 1 (red), 2 (green) and 5 (blue) TeV. {\it Left plots}: $p_T$ of the lepton, $p_T$ of the leading jet and invariant mass of the three leading jets, $M_{Jets}$. {\it Right plots}: pseudorapidity of the third-leading jet, pseudorapidity separation between the third- and the second-leading jet and between the third-leading jet and the lepton.   }
\label{fig:distributions}
\end{center}
\end{figure}

The signal kinematics is typical of a VBF process and is characterized by two final forward-backward jets emitted at high rapidity. The produced $W$ and pseudoscalar have large transverse momenta and their decays produce large missing energy, a large-$p_T$ lepton and a hard $p_T$-leading fat-jet.  
Figure \ref{fig:distributions} shows angular, invariant mass, and $p_T$ distributions for the signal with the $\eta$, after the acceptance cuts. Similar distributions are obtained for the signal with the $\eta^{\prime}$ decaying to $t \bar{t}$. These distributions do not depend on the specific values of the parameters $\sin\theta$ and $d(R)$, which just control the size of the cross section.

\subsection{The background}
The main background comes from $W+\text{jets}$ events. The QCD is the dominant component of this background. After acceptance cuts, the cross section for VBF $W+\text{jets}$ is about 200 fb, more than one order of magnitude smaller than the QCD component. The VBF background is then reduced to a negligible level after the selection criteria we describe in the next section. The background kinematics and angular distributions are indicated with dotted lines in Fig.~\ref{fig:distributions}. Contributions from $t\bar{t}$ and single-top are much smaller after acceptance cuts, with cross sections respectively of 10 fb and 22 fb, and become negligible after applying the signal selection. Likewise, we find that backgrounds from processes with non-topological $\eta/\eta^{\prime}$ interactions are negligible. 
More precisely, if $\eta/\eta^{\prime}$ interacts with the top, we can have a relevant contribution to the final state in (\ref{eq:final-state}) from the top associated production. We find that if the lepton and the missing energy in the final state come from one of the $W$'s emitted from the top, they are much softer than those in our signal and, after the acceptance cuts, the contribution from this process is reduced to a negligible level. On the other hand, if the lepton and the neutrino are produced from the $\eta/\eta^{\prime}$ decay, they are harder and pass the acceptance cuts, but we do not reconstruct correctly the pseudoscalar invariant mass so that, even in this case, this background does not affect our results. Similarly, the process where $\eta/\eta^{\prime}$ is produced by top-mediated gluon fusion can be neglected because its topology is completely different from that of the signal. In fact, after the acceptance cuts in eq. \eqref{eq:acceptanceCuts} we obtain a very small (less than 10$^{-3}$ fb) cross section for the process, which becomes completely negligible after imposing the angular cuts in eq. \eqref{eq:angularCuts}. Here we assumed that the $\eta^{\prime}$ is coupled to the top as in eq. \eqref{eq:h-top}. Another possible contribution to the final state in (\ref{eq:final-state}) comes from the $\eta$ interactions in $\mathcal{L}_{\rm kin}$. In fact, the $\eta$ can be doubly-produced via VBF and the lepton plus missing energy in the final state may originate from one of the two $\eta$'s decaying into $W^{+}W^{-}$. After the acceptance cuts, the cross section for this process is at least one order of magnitude smaller than the typical cross sections of the signal.

To summarize, we can safely neglect in the analysis the background coming from non-topological processes involving the $\eta/\eta^{\prime}$.  

\subsection{The search strategy}

\begin{figure}[t!]
\centering
\includegraphics[width=0.6\textwidth]{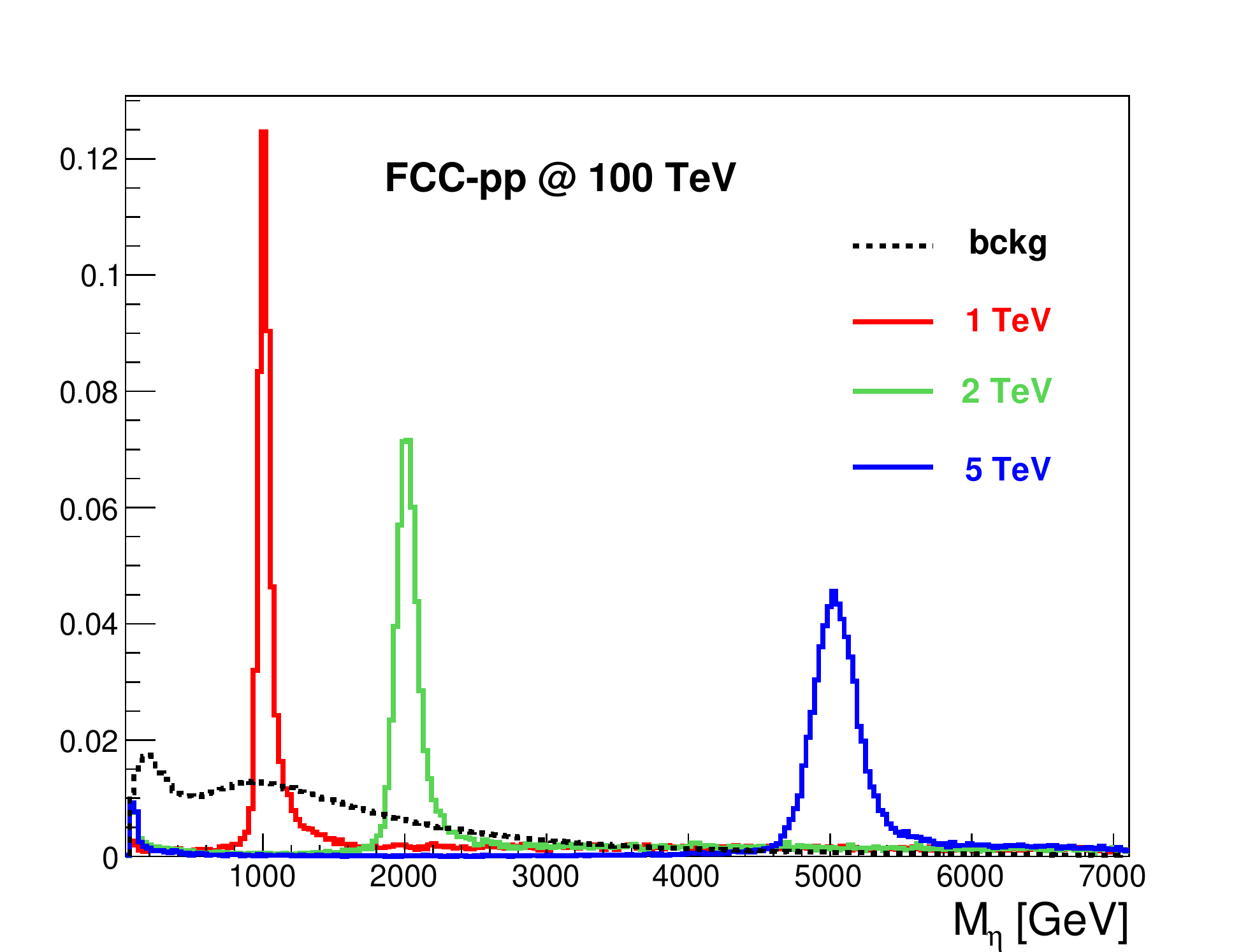}
\caption{\small Distribution normalized to unit area of the invariant mass of the reconstructed $\eta$ particle for background (dotted-black curve) and signal with different $\eta$ mass values of 1 (red), 2 (green) and 5 (blue) TeV. The reconstruction procedure is described in the text.}
\label{fig:Meta}
\end{figure}

We implement the composite Higgs model introduced in section \ref{sec:model} with the WZW interactions detailed in section \ref{sec:wzw-terms} in MG5\_aMC@NLO \cite{Alwall:2014hca}, using FeynRules \cite{Christensen:2008py}.
We simulate both the signal and background with MG5\_aMC@NLO and pass the events to PYTHIA 6.4 \cite{Sjostrand:2006za} for showering and hadronization. 
  In order to mimic detector effects we also apply a Gaussian smearing to the jet energy with:\footnote{Except for the smearing of the jet energies, we do not include systematic uncertainties in our analysis. The impact of systematic errors at the FCC-pp has been considered in e.g. \cite{Cohen:2014hxa,Berlin:2015aba,Alves:2014cda}.}
\begin{equation} \label{eq:smearing}
\frac{\sigma(E)}{E} = C + \frac{N}{E} + \frac{S}{\sqrt{E}} \, \, ,
\end{equation}
where $E$ is in GeV and $C=0.025$, $N=1.7$, $S=0.58$ \cite{Kulchitsky:2000gg}. The jet momentum is then rescaled by a factor $E^{smeared}/E$. Jets are reconstructed with FastJet \cite{Cacciari:2011ma}, applying the Anti-k(t) jet clustering algorithm \cite{Cacciari:2008gp}. In our analysis we will implement a simple reconstruction procedure to identify the pseudoscalar particle using a relatively large cone size $R=$1.5 in the jet clustering. In this way, the hadronic pseudoscalar decay products are collected in a single fat-jet for the majority of signal events. The fat-jet emitted from the decay of the pseudoscalar is the $p_T$-leading jet, $j(1)$. Therefore, the pseudoscalar particle can be simply reconstructed by identifying it with the highest-$p_T$ jet in the final state. To refine the procedure, in the case where the second-leading jet, $j(2)$, is located in the central region, $|\eta_{j(2)}|<2.5$, we reconstruct the pseudoscalar momentum by combining the momenta of the leading and of the second-leading jets. This improves the signal reconstruction for more massive pseudoscalars (above 2 TeV), whose hadronic decays split into two fat-jets for a non-negligible fraction of the events.
Figure \ref{fig:Meta} shows the invariant mass of the reconstructed $\eta$ particle for background and signal with different $\eta$ mass values, after the acceptance cuts. This simple analysis could be refined and improved by applying jet-substructure techniques \cite{Altheimer:2013yza}.\footnote{$ R=1.5 $ is the cone size value for which we obtain the best reconstruction for the $ \eta / \eta' $ resonances. For $ R < 1.5 $ the invariant mass of the reconstructed $ \eta / \eta' $ would present a significant peak in correspondence with the $ W/Z $ mass that would spoil the efficiency of our signal selection, especially for $ \eta / \eta' $ masses above 2 TeV.} Our results for the FCC-pp reach are thus a first conservative estimate.

Based on the characteristic kinematics of the signal, we apply the following selection cuts.\footnote{Note that the cuts implemented in this analysis are chosen after performing a scan over several values and we choose the ones maximizing statistical significance.} We require:
\begin{equation} \label{eq:angularCuts}
|\eta_{j(3)}|> 2 , \quad  |\Delta  \eta \, (j(3) - \ell)|>2 , \quad |\Delta \eta \, (j(3) -j(1))|>2 \, ,
\end{equation}
 where $j(1)$ and $j(3)$ denote the leading and the 3rd-leading jet in $p_T$. As expected from the VBF topology (see Fig. \ref{fig:distributions}), for signal events $j(3)$ is typically emitted at large rapidity and with a large rapidity separation from the lepton and $j(1)$, which are typically located in the central region. Furthermore, depending on the pseudoscalar mass, we apply the cuts listed in Table \ref{tab:cuts}.

 \begin{table}
 \centering
 \begin{tabular}{c| c c c c c}
 $m_{\eta/\eta^{\prime}}$ [TeV] \, & \, $p_T$ $\ell$ [TeV] \, & $\slashed{E}_T$ [TeV] \, & $p_T$ j(1) [TeV] \, & $M_{Jets}$ [TeV] \, & $|\Delta\eta (j(3) -j(2))|$ \\
 \hline 
 && && & \\[-0.3cm]
  0.5 & 0.3 & 0.3 & 4 & 6 & 5 \\
 1 & 0.6 & 0.6 & 4 & 8 & 5 \\
  2 & 0.6 & 0.6 & 5 & 9 & 5 \\
   5 & 1.5 & 1.5 & 6 & 10 & 3   
 \end{tabular}
 \caption{\small  Selection cuts for different pseudoscalar mass values. $M_{Jets}$ indicates the invariant mass of the first three leading jets in $p_T$.}
 \label{tab:cuts}
 \end{table}
Finally, we apply a cut on the invariant mass of the reconstructed pseudoscalar ($M_{\eta/\eta^{\prime}}$):
\begin{equation}
\begin{tabular}{c | c c c c }
$m_{\eta/\eta^{\prime}}$ [TeV] & 0.5 & 1 & 2 & 5  \\
\hline
 && &&  \\[-0.3cm]
$M_{\eta/\eta^{\prime}}$ [TeV] & [0.48, 0.54] & [0.93, 1.2] & [1.9, 2.2] & [4.6, 5.8] 
\end{tabular}
\end{equation}

\subsection{Results}

 In table \ref{tab:eta-xsec} we report the cross sections for the background and the signal with an $\eta$ and with an $\eta^{\prime}$ at different masses and for the reference values: $\sin\theta = 0.25$ and $d(R)=10$. We apply the same selection for the signal with the $\eta$ and with the $\eta^{\prime}$.
 From our results we can calculate the FCC-pp reach on our signal process as a function of $m_{\eta \slash \eta^{\prime}}$, $\sin\theta$ and $d(R)$. The signal cross section depends quadratically on $d(R)$ while, to a very good approximation, it is proportional to $\sin^2 \theta \cos^2 \theta$, for the process with the $\eta$, and to $\sin^2 \theta$ for the $\eta^{\prime}$ signal, see eqs. \eqref{eq:WZW-AAeta}, \eqref{eq:WZW-AAetap} and \eqref{eq:WZW-AAAeta}. The statistical significance is estimated according to
\begin{equation}
\frac{N_S}{\sqrt{N_S + N_B}} \ ,
\end{equation}
where $N_S$ ($N_B$) is the number of signal (background) events. We claim a 5$\sigma$ discovery when this ratio is equal to 5, while we obtain an estimate of the 95\% CL exclusion reach when it is 2.

   \begin{table}
   \begin{center}
   \begin{tabular}{c c| c c c c }
   $m_{\eta \slash \eta^{\prime}}$ [TeV] & & 0.5 & 1 & 2 & 5  \\
   \hline
 && && &  \\[-0.3cm]
   Bckg & Accept. & \multicolumn{4}{c}{6.6 $\cdot$ 10$^{3}$} \\
   & Final & 0.056 & 0.040 & 0.017 & 0.0096  \\
    \hline
 && && &  \\[-0.3cm]
 S ($\eta$) & Accept. & 0.15 & 0.17 & 0.16 & 0.048   \\
 & Final & 0.019 & 0.029 & 0.023 & 0.0072   \\
  && && &  \\[-0.3cm]
 S ($\eta^{\prime}$) & Accept. & 0.020 & 0.021 & 0.018 & 0.011 \\
 & Final & 0.0026 & 0.0040 & 0.0023 & 0.0010 
   \end{tabular}
   \caption{\small Cross section values (in fb) at the FCC-pp with $\sqrt{s}=100$ TeV for the background (Bckg) and the signal with an $\eta$ (S$(\eta$)) and with an $\eta^{\prime}$ (S$(\eta^{\prime})$) at different masses, after the acceptance cuts and after the complete selection. We fix $\sin\theta = 0.25$ and $d(R)=10$.}
   \label{tab:eta-xsec}
   \end{center}
   \end{table}

Figure \ref{fig:reach-eta} shows the resulting reach of the FCC-pp on topological interactions involving an $\eta$ and an $\eta^{\prime}$. We indicate the discovery and exclusion reach for a pseudoscalar of 1 TeV on the plane ($d(R)$, $\sin\theta$) and the 95\% CL exclusion curves for different $\sin\theta$ values on the plane ($m_{\eta/\eta^{\prime}}$, $d(R)$). We get promising results for the FCC-pp. With an $\eta$ of 1 TeV and a value $\sin\theta \approx 0.2$, the FCC-pp with 3 ab$^{-1}$ can test the underlying strong dynamics for $d(R)$ as small as 7, whereas a discovery can occur for $d(R)\gtrsim 11$. Considering the large-$N$ estimate in \eqref{eq:g-rho}, this corresponds to test an effective coupling among composite resonances $g_{\rho}\lesssim 5$. Notice that the reach remains high even for a heavier $\eta$, i.e. it does not decrease significantly up to $m_{\eta} \approx$ 5 TeV.  The FCC-pp reach on the signal with the $\eta^{\prime}$ is slightly smaller due to a reduced production cross section, as shown in Fig. \ref{fig:xsec}.  

\section{Summary and outlook}\label{sec:conclusion}

In this paper, we have studied the phenomenology resulting from the topological sector of a composite Higgs theory defined in the coset $\SU(4)/\Sp(4)$ with underlying fundamental fermionic matter, including new pseudoscalar composite states. We computed the relevant WZW anomalous interactions of the theory and found new promising production mechanisms for the composite spin-0 states, in particular via VBF in association with a $W/Z$ or a Higgs boson. These processes represent the dominant $\eta/\eta^{\prime}$ production channels at a future 100 TeV proton-proton collider, FCC-pp. Indeed, our search strategy  provides a promising discovery channel for the $ \eta/\eta' $ resonances. We remark that the calculation of these processes relies on an expansion of the WZW action up to dimension-6 operators, which have been neglected in previous studies. 

\begin{figure}[t!]
\begin{center}
\includegraphics[width=0.48\textwidth, height=5.3cm]{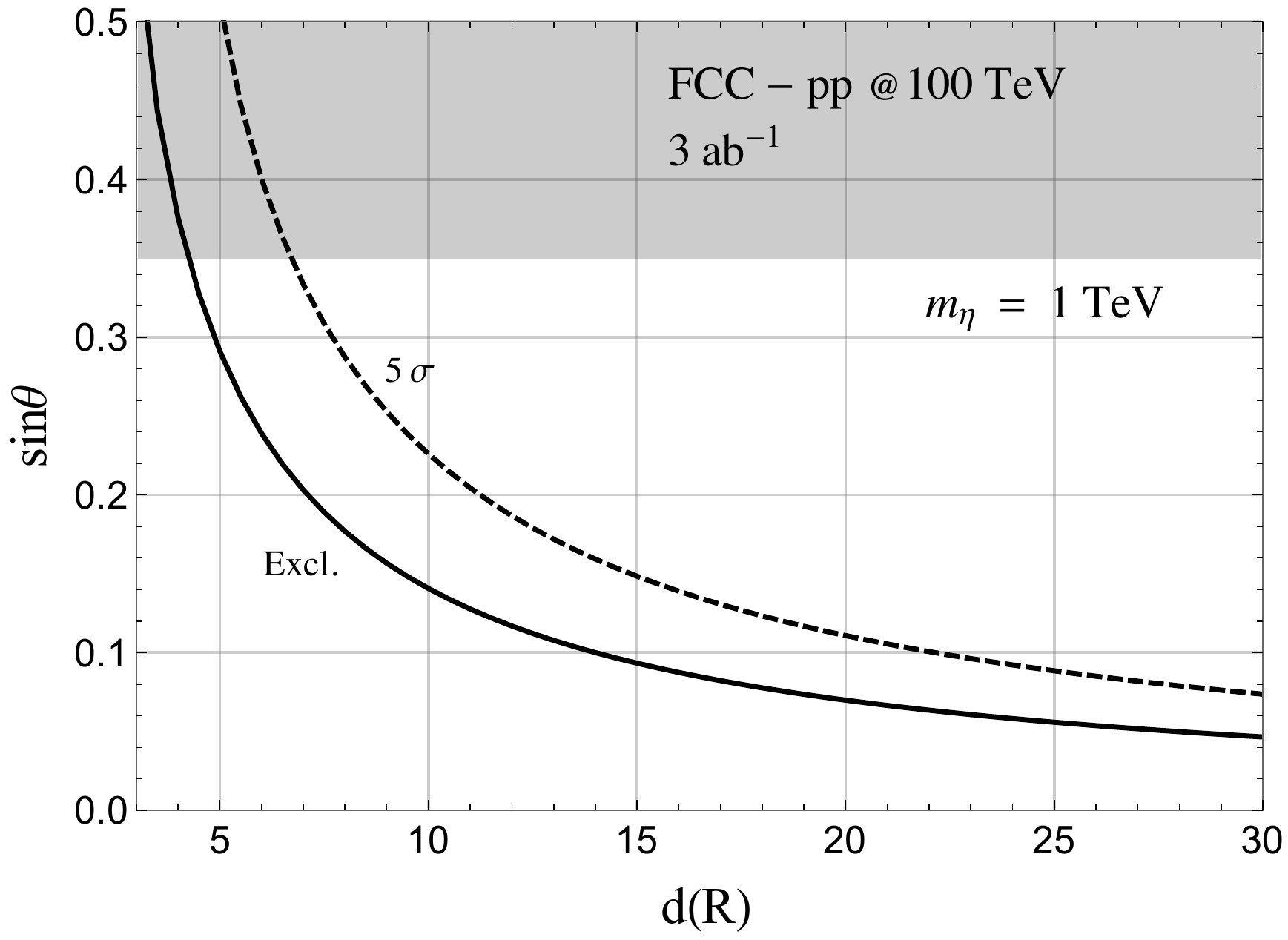}\quad
\includegraphics[width=0.48\textwidth, height=5.3cm]{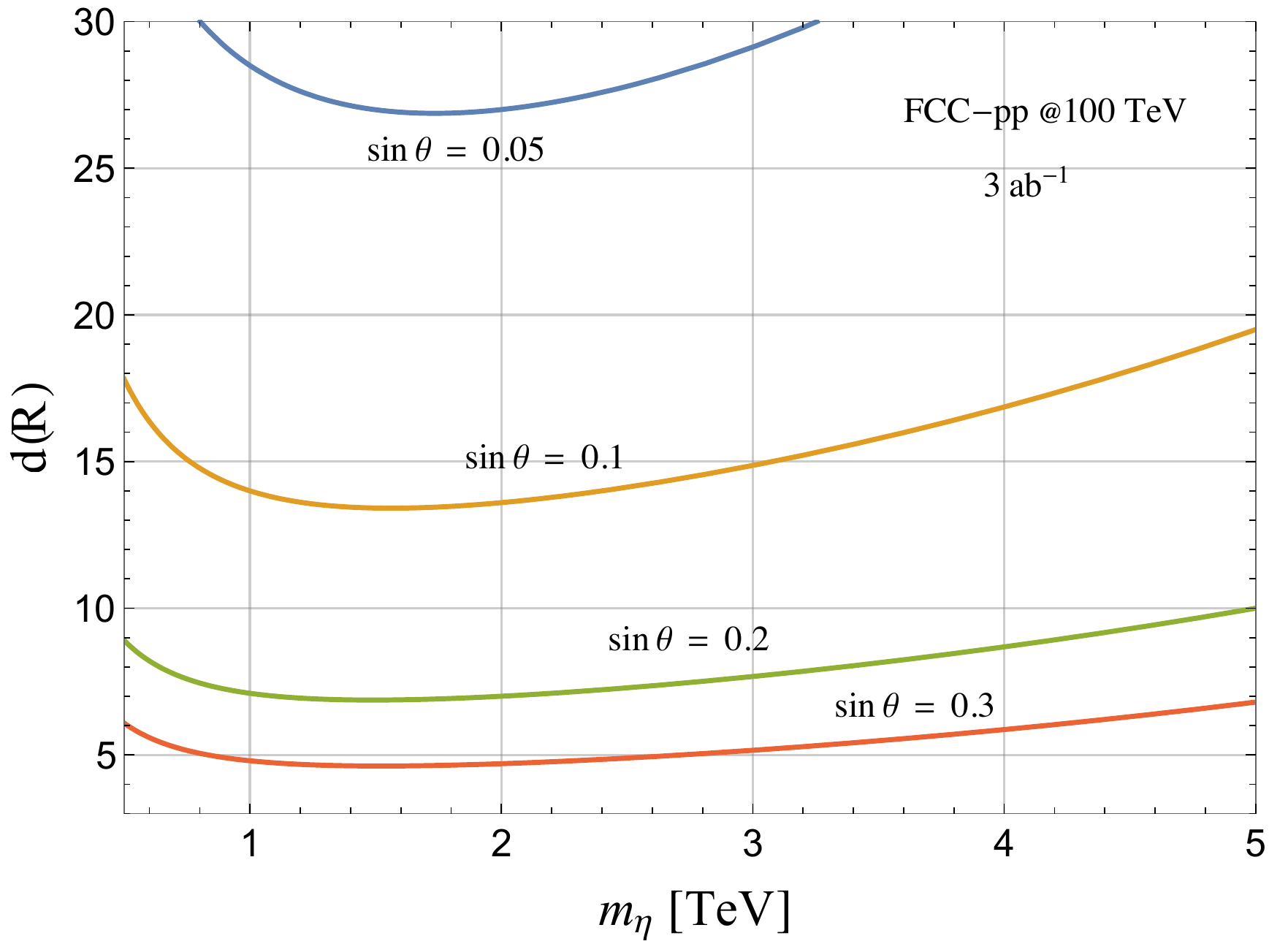} \\ \vspace{0.25cm}
\includegraphics[width=0.48\textwidth, height=5.3cm]{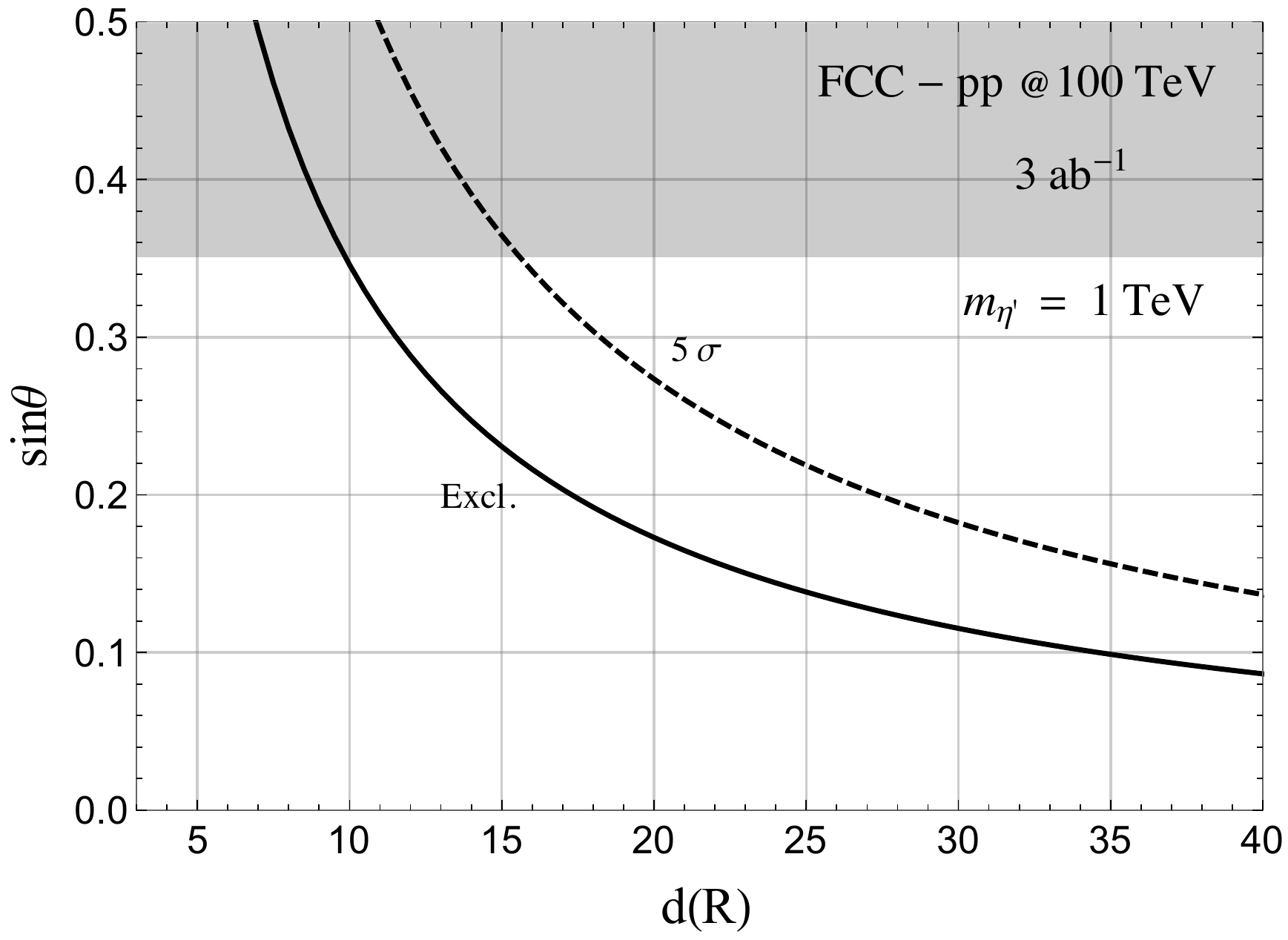}\quad
\includegraphics[width=0.48\textwidth, height=5.3cm]{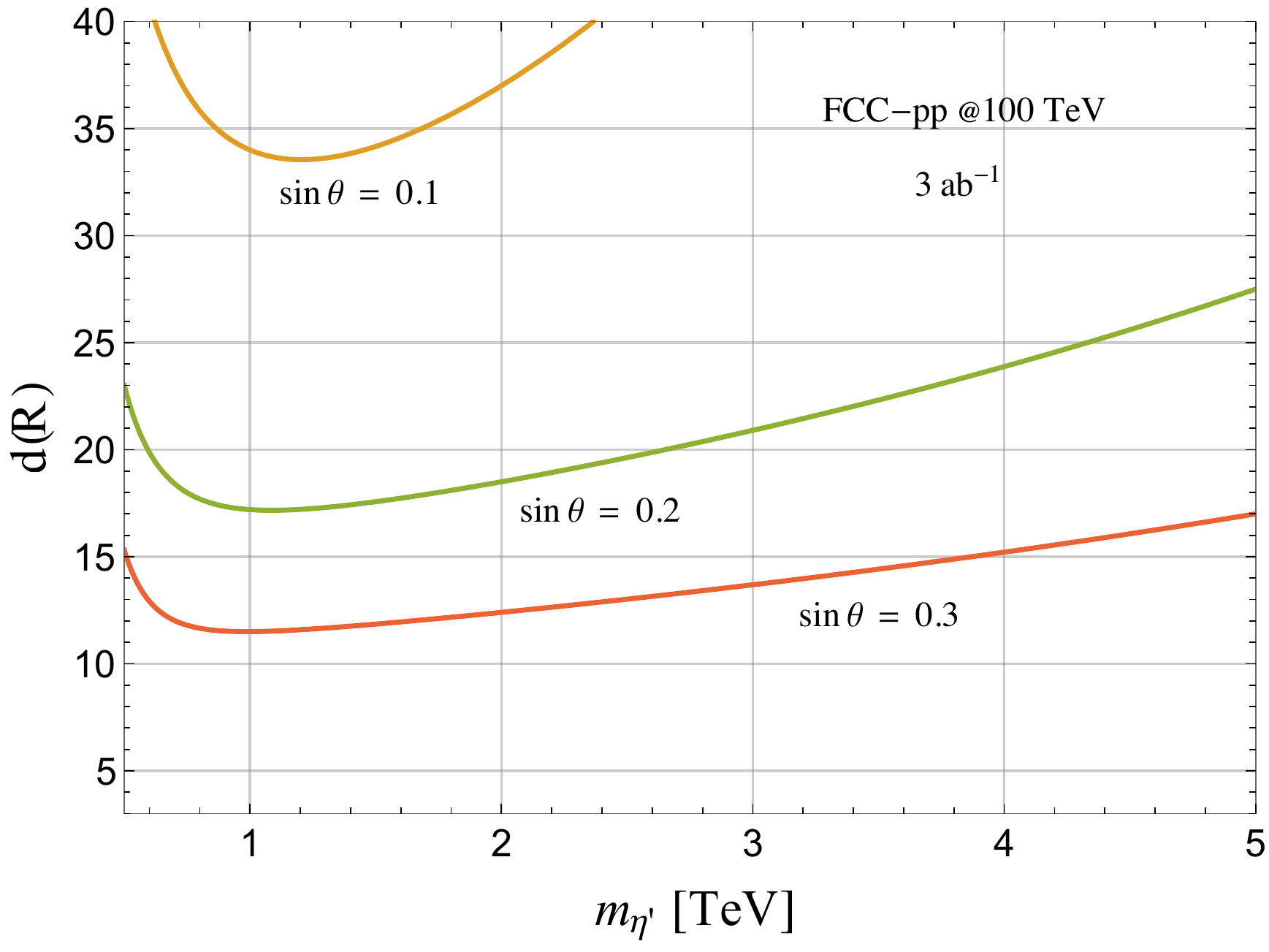}
\caption{ \small Reach at a 100 TeV proton-proton collider with a luminosity of 3 ab$^{-1}$. {\it Upper panels (lower panels)}: results for the signal with an $\eta$ ($\eta^{\prime}$). {\it Left panels}: 5$\sigma$ (dashed curve) discovery and 95\% CL exclusion (continuous line) reach for the signal with $m_{\eta/\eta^{\prime}}=$1 TeV in the plane ($d(R)$, $\sin\theta$). The shaded area is excluded by LHC Higgs data \eqref{eq:h-stheta} if the Higgs Yukawa coupling reads as in eq. \eqref{eq:h-top}. {\it Right panels}: 95\% CL exclusion curves for different $\sin\theta$ values on the plane ($m_{\eta/\eta^{\prime}}$, $d(R)$). }
\label{fig:reach-eta}
\end{center}
\end{figure}

We performed a detailed analysis for the channel $\eta/\eta^{\prime}+ W$, showing that this process is able to provide direct information on the fundamental structure of the composite Higgs theory, namely the number of degrees of freedom characterizing the new strong sector, $d(R)$, and the vacuum alignment angle $\theta$. Our final results are summarized in Fig.~\ref{fig:reach-eta}, where it is shown that the FCC-pp can test the underlying gauge dynamics for $d(R)$ as small as 7 (5) and $\sin\theta=0.2 \, (0.3)$.

 Our study points out the importance of new channels with topological interactions to test strong dynamics beyond the SM, related to the mechanism of EW symmetry breaking. We considered a minimal realization with just two scalar resonances in addition to the composite Higgs and without vector-like top partners. This analysis can also be generalized to scenarios with larger cosets and particle content. 

\section*{Acknowledgments}
The CP$^{3}$-Origins center is partially funded by the Danish National Research Foundation, grant number DNRF90.

\newpage
\appendix

\mathversion{bold}
\section{The broken $ \SU(4) $ generators} \label{app:generators}
\mathversion{normal}

We report the broken generators of the $ \SU(4) $ flavor symmetry, due to the formation of a condensate with the alignment given in eq. \eqref{eq:vacuum}:
	\begin{equation}
	\begin{split}
 Y^1 = \dfrac{1}{2\sqrt{2}}& \begin{pmatrix} -s_\theta \sigma^1 & c_\theta \sigma^3 \\ c_\theta \sigma^3 & -s_\theta \sigma^1 \end{pmatrix}, \quad 
	Y^2 = \dfrac{1}{2\sqrt{2}} \begin{pmatrix} s_\theta \sigma^2 & i c_\theta \mathbf{1} \\ - i c_\theta \mathbf{1} & -s_\theta \sigma^2 \end{pmatrix}, \\
	Y^3 = \dfrac{1}{2\sqrt{2}} \begin{pmatrix} s_\theta \sigma^3 & c_\theta \sigma^1 \\ c_\theta \sigma^1 & s_\theta \sigma^3 \end{pmatrix}, 
	&\quad Y^4 = \dfrac{1}{2\sqrt{2}} \begin{pmatrix} 0 & \sigma^2 \\ \sigma^2 & 0 \end{pmatrix}, \quad
	Y^5 = \dfrac{1}{2\sqrt{2}} \begin{pmatrix} c_\theta \mathbf{1} & -i s_\theta \sigma^2 \\ i s_\theta \sigma^2 & -c_\theta \mathbf{1} \end{pmatrix}.
	\end{split}
	\end{equation}
We have introduced the shorthand notation: $ c_\theta =\cos \theta $ and $ s_\theta = \sin \theta $.

\section{Wess-Zumino-Witten action}\label{app:WZW}
For completeness, we refer the full form of the gauged WZW action for a model with the pNGB states described by a non-linear $ \Sigma $-model in the coset $\SU(4)/\Sp(4)$ ~\cite{Duan:2000dy}.\footnote{We corrected a factor two in the normalization of the overall coefficient $ c $ reported in \cite{Duan:2000dy}.}~Using the notation introduced in eq. \eqref{eq:WZWnotation}, the action reads:
	\begin{equation}
	\begin{split}
	\Gamma_\WZW =\, & c \int_{M^5} \Tr\left[\alpha^5 \right] + 10 i \,c \int_{M^4} \Tr \left[\mathcal{A} \alpha^3 \right] \\
	&- 10 c  \int_{M^4} \Tr \left[ (\dd \mathcal{A}\,  \mathcal{A} + \mathcal{A} \dd \mathcal{A}) \alpha \right] - 5 c \int_{M^4} \Tr \left[\dd \mathcal{A} \dd \Sigma \mathcal{A}\transpose \Sigma^{\dagger} - \dd \mathcal{A}\transpose \dd \Sigma^\dagger \mathcal{A} \Sigma \right] \\
	&-5 c\int_{M^4} \Tr \left[ \Sigma \mathcal{A}\transpose \Sigma^\dagger(\mathcal{A} \alpha^2 + \alpha^2 \mathcal{A}) \right] +5 c \int_{M^4} \Tr \left[ (\mathcal{A}\alpha)^2 \right] \\
	& + 10 i \, c \int_{M^4} \Tr \left[ \mathcal{A}^3 \alpha \right] + 10 i \, c \int_{M^4} \Tr \left[ (\dd \mathcal{A}\, \mathcal{A} + \mathcal{A}\, \dd \mathcal{A}) \Sigma \mathcal{A}\transpose \Sigma^\dagger \right] \\
	&- 10 i \, c \int_{M^4} \Tr \left[ \mathcal{A} \alpha \mathcal{A} \Sigma \mathcal{A}\transpose \Sigma^\dagger \right] \\
	&+ 10 c \int_{M^4} \Tr \left[ \mathcal{A}^3 \Sigma \mathcal{A}\transpose \Sigma^\dagger  \right] + \tfrac{5}{2} c \int_{M^4} \Tr\left[ (\mathcal{A} \Sigma \mathcal{A}\transpose \Sigma^\dagger)^2 \right], 
	\end{split}
	\end{equation}
	with $ c= -i d(R)/480\pi^2 $.
The first term is the usual Wess-Zumino action \cite{Wess:1971yu}, which is given in five-dimensional space. The remaining terms are constructed so that they cancel any gauge variation term by term.

\mathversion{bold}
\section{Decay rates for the $\eta$ and the $\eta^{\prime}$} \label{app:rates}
\mathversion{normal}

We report below the (2-body) decay rates for the $\eta$ resulting from the WZW action: 

\begin{align}
\begin{split}
\Gamma{(\eta \to \gamma Z)} & = d(R)^2 \sin^2\theta \cos^2\theta  \frac{\alpha^2_{\rm em}}{512 \pi^3 v^2}\frac{1}{c^2_w s^2_w} \, m^3_{\eta} \, , \\
\Gamma{(\eta \to Z Z)} & = d(R)^2 \sin^2\theta \cos^2\theta  \frac{\alpha^2_{\rm em}}{1024 \pi^3 v^2}\frac{(1-2 s^2_w)^2}{c^4_w s^4_w} \, m^3_{\eta} \, ,\\
\Gamma{(\eta \to W^+ W^-)} & = d(R)^2 \sin^2\theta \cos^2\theta  \frac{\alpha^2_{\rm em}}{512 \pi^3 v^2}\frac{1}{ s^4_w} \, m^3_{\eta} \, ,
\end{split}
\end{align}
and for the $\eta^{\prime}$, from the coupling to the top in \eqref{eq:h-top}:

\begin{align}
\begin{split}
\Gamma{(\eta^{\prime} \to t\bar{t})} & =  \sin^2\theta  \frac{3 \, m^2_{top}}{8 \pi v^2} \, m_{\eta^{\prime}} \sqrt{1 -4 \frac{m^2_{top}}{m^2_{\eta}}} \, ,\\
\Gamma{(\eta^{\prime} \to \gamma \gamma)} & = d(R)^2 \sin^2\theta  \frac{\alpha^2_{\rm em}}{256 \pi^3 v^2} \, m^3_{\eta^{\prime}}\, , \\
\Gamma{(\eta^{\prime} \to \gamma Z)} & = d(R)^2 \sin^2\theta  \frac{\alpha^2_{\rm em}}{512 \pi^3 v^2}\frac{(1-2s^2_w)^2}{c^2_w s^2_w} \, m^3_{\eta^{\prime}} \, , \\
\Gamma{(\eta^{\prime} \to Z Z)} & = d(R)^2 \sin^2\theta   \frac{\alpha^2_{\rm em}}{9216 \pi^3 v^2}\frac{(3-6 s^2_w c^2_w-\sin^2\theta)^2}{ c^4_w s^4_w} \, m^3_{\eta^{\prime}}\, , \\
\Gamma{(\eta^{\prime} \to W^+ W^-)} & = d(R)^2 \sin^2\theta   \frac{\alpha^2_{\rm em}}{4608 \pi^3 v^2}\frac{(3-\sin^2\theta)^2}{ s^4_w} \, m^3_{\eta^{\prime}} \, .
\end{split}
\end{align}
For very massive pseudoscalars, $m_{\eta/\eta^{\prime}}> 5$ TeV, the 3-body decays into gauge bosons mediated by 3 and 4-boson topological interactions become dominant over the 2-body decays.

\bibliography{WZW.bib}

\end{document}